\documentclass[aps,twocolumn,nopacs,nofootinbib,floatfix,superscriptaddress]{revtex4}
\usepackage{graphicx}
\usepackage{amsfonts}
\usepackage{amssymb}
\usepackage{amsbsy}
\usepackage{amsmath}
\usepackage{mathrsfs}
\usepackage{latexsym}
\usepackage{natbib}
\usepackage{bm}
\usepackage{subfigure} 
\usepackage{color}
\usepackage{wasysym}
\usepackage{mathbbol}
\usepackage{bigints}
\allowdisplaybreaks
\usepackage[normalem]{ulem}
\usepackage[dvipsnames]{xcolor}
\usepackage{multirow}
\usepackage{csquotes}

\RequirePackage[colorlinks,citecolor=blue,urlcolor=magenta,linkcolor=blue]{hyperref}
\allowdisplaybreaks

\def\gr{general relativity}

\def\gr{general relativity}

\labelformat{chapter}{Chapter (#1)} 
\labelformat{section}{Sec. #1} 
\labelformat{subsection}{Sec. #1} 
\labelformat{subsubsection}{Sec. #1}
\labelformat{subsubsubsection}{Sec. #1)}
\labelformat{equation}{Eq.~(#1)} 
\labelformat{figure}{Fig.~#1} 
\labelformat{subfigure}{Fig.~(\thefigure#1)} 
\labelformat{table}{Tab.~(#1)} 
\labelformat{appendix}{Appendix #1}

\def\gr{general relativity}

\def \GB {Gauss-Bonnet}

\def\gr{general relativity}

\def \GB {Gauss-Bonnet}

\begin{document}
\title{Pure Gauss-Bonnet NUT Black Hole Solution: II}

\author{Sajal Mukherjee}
 \email{sajal.mukherjee@pilani.bits-pilani.ac.in}
 \affiliation{Birla Institute of Technology and Science Pilani, Rajasthan, 333031, India}

\author{Subhajit Barman}
\email{subhajit.barman@physics.iitm.ac.in}
\affiliation{Centre for Strings, Gravitation and Cosmology, Department of Physics, Indian Institute of Technology Madras, Chennai 600036, India}

\begin{abstract}
\noindent
In the present article, we have obtained an exact analytical solution of six-dimensional pure Gauss-Bonnet gravity in the presence of both NUT and Maxwell charges. The topology of the horizon is chosen to be the product of two 2-spheres. Upon evaluating the solution, we study the spacetime properties, such as the event horizon and singularity, and obtain the ranges of parameter space where the solution is valid. We discuss how the presence of Maxwell charges may impact the solution's asymptotic expansion and what distinctive effects it will bring to the geometry. The thermodynamic properties of the solution are also discussed, emphasizing the interplay between NUT and Maxwell charges.
\end{abstract}

\maketitle

\section{Introduction}
Due to the advancement in observational aspects of gravity,  test of general relativity (GR) has become a matter of great concern in recent times \cite{LIGOScientific:2018mvr, LIGOScientific:2019fpa, LIGOScientific:2020tif,EventHorizonTelescope:2019dse,EventHorizonTelescope:2019uob}. By inferring the observational findings, it is, in principle, possible to constrain the deviation from GR \cite{Garcia_Quintero_2020,LIGOScientific:2021sio}. Over the years, several alternatives to GR have been developed, namely $f(R)$ \cite{Sotiriou:2008rp}, Gauss-Bonnet \cite{Boulware:1985wk}, scalar-tensor theories \cite{Brans:2008zz,Nordtvedt:1970uv,Wagoner:1970vr}, Lovelock \cite{Lovelock:1971yv}, etc. In addition to the interest in testing these theories from observation, it is also important to obtain black hole (BH) solutions in these theories. In many cases, it has become challenging to obtain black hole (BH) solutions due to the complexity of the field equations. In the present paper, we aim to find an analytical solution for one of such alternatives to GR, namely, Gauss-Bonnet gravity. We have a similar attempt in Ref. \cite{Mukherjee:2021erg} (hereafter paper-I), where an exact solution in Gauss-Bonnet gravity is derived.

\GB~theory of gravity is a part of a more general theory known as the Lovelock theory. Due to its interesting properties, the Lovelock theory has gained popularity in the community, and various aspects of Lovelock have been explored in the past \cite{Myers:1988ze, Wheeler:1985nh, Banados:1992wn,Dadhich:2012ma}. It is known to be a natural generalization of the general theory of relativity in higher dimensions. Despite having higher order corrections in the action, the field equations remain second order and free of ghosts \cite{Zwiebach:1985uq,Deruelle:1989fj}. In general, the action in the Lovelock theory is composed of contributions from different orders. For example, the 3rd-order Lovelock action may be written as the sum of first-order (GR), second-order (\GB), and third-order action. However, one is also left with a choice to pick individual contributions from different orders by ignoring the sum over all lower orders \cite{Camanho:2015hea}. This way, one may expect to obtain a completely new solution without any general relativistic imprints. We call them pure Lovelock solutions \cite{Dadhich:2015nua}, and we will consider one of such solutions in this study. These solutions carry several useful theoretical implications, such as kinematic in critical odd dimension for N-th order, $d=2N+1$ \cite{Dadhich:2012cv}, the existence of bound orbits \cite{Dadhich:2013moa}, and their stability \cite{Gannouji:2019gnb}. In particular, we are interested in the $N=2$ case, i.e., pure \GB~gravity.

The other pillar of the present paper is associated with NUT charge \cite{Newman:1963yy,Bueno:2018uoy}. Besides introducing additional features \cite{Bordo:2019tyh,Hennigar:2019ive}, NUT BH adds nicely to the Kerr-Newman family of BH solutions. It breaks down the asymptotic flatness of the solution, which saves Birkhoff's theorem \cite{d1992introducing} in the static limit and makes NUT BH a consistent solution of the vacuum Einstein field equations. The theoretical implication of its origin has been studied in detail \cite{LyndenBell:1996GM}. For example, it is argued that the NUT charge can be connected with a gravomagnetic field and may act as a gravomagnetic charge \cite{Turakulov:2001bm}. In addition to the theoretical implications, in recent years, there have also been some observational avenues that relate to NUT charge. For example, the NUT charge changes the multipole structure of the spacetime, which may have some nontrivial imprints on the emitted GWs \cite{Mukherjee:2020how}. The other observation may be associated with X-ray binary \cite{Chakraborty:2017nfu} and the non-existence of equatorial circular orbits \cite{Datta:2020axm}. Another aspect that we should touch upon is the horizon topology. In higher-dimensional spacetime, theoretically, one may assign different horizon topologies. For example, a 6-dimensional static spacetime can have spherical $S^{(4)}$, or $S^{(1)} \times S^{(3)}$, or product of two 2-spheres topology $S^{(2)} \times S^{(2)}$. In paper-I, we have explored the product topology, and in the present paper too, we will continue to consider that. The reason is the presence of the NUT charge, which introduces a cross-term, and an analytical solution with spherical topology becomes unlikely. 
At this point, we should also comment on the chosen dimensionality of the problem in the present context, that is six. Note that to find a NUT BH solution in higher dimensions, the base space needs to be K$\ddot{\text{a}}$hler-Einstein manifold \cite{Bueno:2018uoy}. Turns out, the product topology of two spheres, i.e., $S^{(2)} \times S^{(2)}$, is one of the simple choices for this to happen. Therefore, a six-dimensional spacetime can be a natural choice to start with. However, it is possible to extend the present approach to higher-dimensional spacetimes, as long as the dimension of the spacetime is even. Within the even spacetime, one can easily incorporate the product topology of the horizon.

In paper-I, our study was limited to finding NUT BH solutions in pure \GB~gravity. While in the present study, we introduce an electromagnetic field and solve the \GB-Maxwell's field equations -- to capture any effects that may emerge from the interplay between Maxwell and NUT charges. In particular, both the near-horizon and asymptotic structure of the solution would carry imprints of these charges and probably encode effects emerging from the coupling between them.
In addition to the classical implications of the considered background, we are interested in exploring semi-classical features, which can give illuminating insight of the spacetime. In this regard, the horizon thermodynamics of the concerned black hole can play a crucial role. It is known that the characteristic temperature corresponding to the Hawking effect contains important information about the black hole parameters. Such as, in static spacetimes, this temperature is higher for smaller mass black holes \cite{Hawking:1975vcx, Birrell:1982ix, Fulling:1989nb, Ford:1997hb, Traschen:1999zr, Jacobson:2003vx, Lambert:2013uaa, Barman:2017fzh, Barman:2017vqx} and it depends nontrivially on the charge \cite{Good:2020qsy, McMaken:2023tft, Barman:2023rhd}. The temperature also depends on the angular momentum of a rotating black hole and the condition of extremality \cite{Barman:2018ina, Barman:2021gcd, Ghosh:2021ijv}. In higher-dimensional spacetimes, the temperature gets affected by the presence of the extra dimensions \cite{Kanti:2004nr, Barman:2019vst} and also the rotation corresponding to different directions \cite{Frolov:2002xf}. Moreover, with rotation in the spacetime, the spectrum of the Hawking effect contains the superradiance effect, where depending on the angular momentum of a certain field mode the spectrum can get amplified. All these observations make it fascinating to study the Hawking effect in our concerned background, as it also contains aspects of similarity with the Kerr-Newman BHs. It will also be interesting to check whether the spectrum and temperature of the Hawking effect can indicate features that can distinguish these types of BH solutions from others.

The rest of the paper is organized as follows. In \ref{sec:Preli} and \ref{sec:EM_part}, we introduce the field equations and electromagnetic field tensor, respectively. Following that, we derive the BH solution in \ref{sec:obtain_metric}, and discuss the validity of the solution in \ref{sec:Validity}. Next, in \ref{sec:thermo}, we have studied the thermodynamic properties of this particular BH solution. Finally, we conclude the paper in \ref{sec:Discussion}.

\textit{Notation and convention:} We have set the constants $c=G=1$, make use of the metric convention $(-,+,+,+,+,+)$, and $(t,r,\theta_1,\phi_1,\theta_2,\phi_2) \equiv (0,1,2,3,4,5)$. For index $0$ to $5$, we use Greek letters, while to denote spatial components running from $1$ to $5$, we use Latin letters.

\section{Field equations in pure Gauss-Bonnet gravity} \label{sec:Preli}
We start with the following action in D dimensions
\begin{equation}
\mathcal{S}=\int d^Dx \sqrt{-g}\bigl(\alpha_2\mathcal{L}_{\rm GB}+\mathcal{L}_{\rm m}-2 \Lambda \bigr),
\end{equation}
where, $g$ is the metric's determinant, $\mathcal{L}_{\rm m}$ is the matter Lagrangian, $\alpha_2$ is the \GB~coupling constant, and $\Lambda$ is the positive cosmological constant. In the above, $\mathcal{L}_{\rm GB}$ is defined as the Gauss-Bonnet Lagrangian with the following expression:
\begin{equation}
\mathcal{L}_{\rm GB}=R^2-4R^{\alpha \beta}R_{\alpha \beta}+R^{\alpha \beta \gamma \delta}R_{\alpha \beta \gamma \delta}.
\label{eq:LGB}
\end{equation}
In the above, $R$, $R_{\alpha \beta}$ and $R_{\alpha \beta \gamma \delta}$ are Ricci scalar, Ricci tensor and Riemann tensor, respectively. For the above action, the field equation reads as
\begin{equation}
H_{\alpha \beta}=\alpha_{2} \left(2 J_{\alpha \beta}-\dfrac{1}{2}g_{\alpha \beta} \mathcal{L}_{\rm GB} \right)=-\Lambda g_{\alpha \beta}+T_{\alpha \beta},
\label{eq:field_equation}
\end{equation}
where $g_{\alpha \beta}$ is the metric tensor, and $T_{\alpha \beta}$ is given as the stress-energy tensor appears due to the Lagrangian $\mathcal{L}_{\rm m}$. In the above expression, $H_{\alpha \beta}$ and $J_{\alpha \beta}$, play the role analogous to Einstein and Ricci tensor in Einstein's gravity \cite{Padmanabhan:2013xyr}. Furthermore, $J_{\alpha \beta}$ can be expanded in terms of Riemann and Ricci as follows:
\begin{equation}
J_{\alpha \beta}=RR_{\alpha \beta}-2R^{\gamma}_{~\alpha}R_{\beta \gamma}-2R^{\gamma \delta}R_{\alpha \gamma \beta \delta}+R_{\alpha}^{~\gamma \delta \kappa}R_{\beta \gamma \delta \kappa}.
\label{eq:Jab}
\end{equation}
As in the present context, we discuss a case in the presence of an electromagnetic field, the following equation will be of particular use:
\begin{equation}
T^{\alpha}_{~\beta}=2\left\{F^{\alpha \gamma}F_{\beta \gamma}-\dfrac{1}{4}\delta^{\alpha}_{\beta}F^{\gamma \delta}F_{\gamma \delta}\right\},
\label{eq:Energy_momentum}
\end{equation}
where $F^{\alpha \beta}$ can be constructed from the vector potential $A^{\alpha}$ as given below
\begin{equation}
F_{\alpha \beta}=\partial_{\alpha}A_{\beta}-\partial_{\beta}A_{\alpha}.
\label{eq:vector_potential}
\end{equation} 
Finally, the electromagnetic field tensor $\mathbf{F}$, a differential 2-form, can be written in terms of the components $F^{\alpha \beta}$ as follows:
\begin{equation}
 \mathbf{F}=F_{\alpha \beta}dx^{\alpha}\wedge dx^{\beta},
\end{equation}
where \enquote*{$\wedge$} is known as the outer or wedge product.

With the primary equations being introduced, we aim to obtain the exact solutions within the pure Gauss-Bonnet gravity in the upcoming sections.
\section{The electromagnetic field tensor} \label{sec:EM_part}
In the source-free region, the components of the Maxwell's field tensor, $F^{\mu \nu}$, satisfy the following divergence-free condition
\begin{equation}
\partial_{\nu}(\sqrt{-g}F^{\mu \nu})=0,
\label{eq:field_eq}
\end{equation}
where $g$ is the determinant of the metric. 
In order to solve the field equations, we must specify the field tensor $\mathbf{F}$. 
Note that the electric ($E^{\alpha}$) and magnetic field ($B^{\alpha}$) simply relates $F^{0\mu}$ and spatial component $F^{ij}$ of the field tensor respectively. In the case when spacetime is spherically symmetric, which naturally inherits the radial symmetry, all the magnetic field components identically vanish. Besides, the radial symmetry will also make the angular components of the electric field zero. Therefore, the field tensor becomes, $\mathbf{F}=F_{01}~dt \wedge dr=E_{1}~dt \wedge dr$. If we now add the NUT charge, the spherical symmetry would be destroyed, but given that the solution is static, the radial symmetry will continue to exist. This means that both $E^{\theta}$ and $E^{\phi}$ would vanish, but the magnetic field components would survive. Due to this reason, the electromagnetic field 2-form is given as \cite{Mann:2005mb}:
%
\begin{equation}
F=\dfrac{Q}{(r^2+l^2)^2}(r^2-l^2)\left\{dr \wedge [dt+2 l \cos\theta d\phi]\right\},
\end{equation}
where $Q$ and $l$ are the electric and NUT charge respectively, and both electric and magnetic fields are non-zero.

\begin{figure}[htp]
 \includegraphics[scale=0.39]{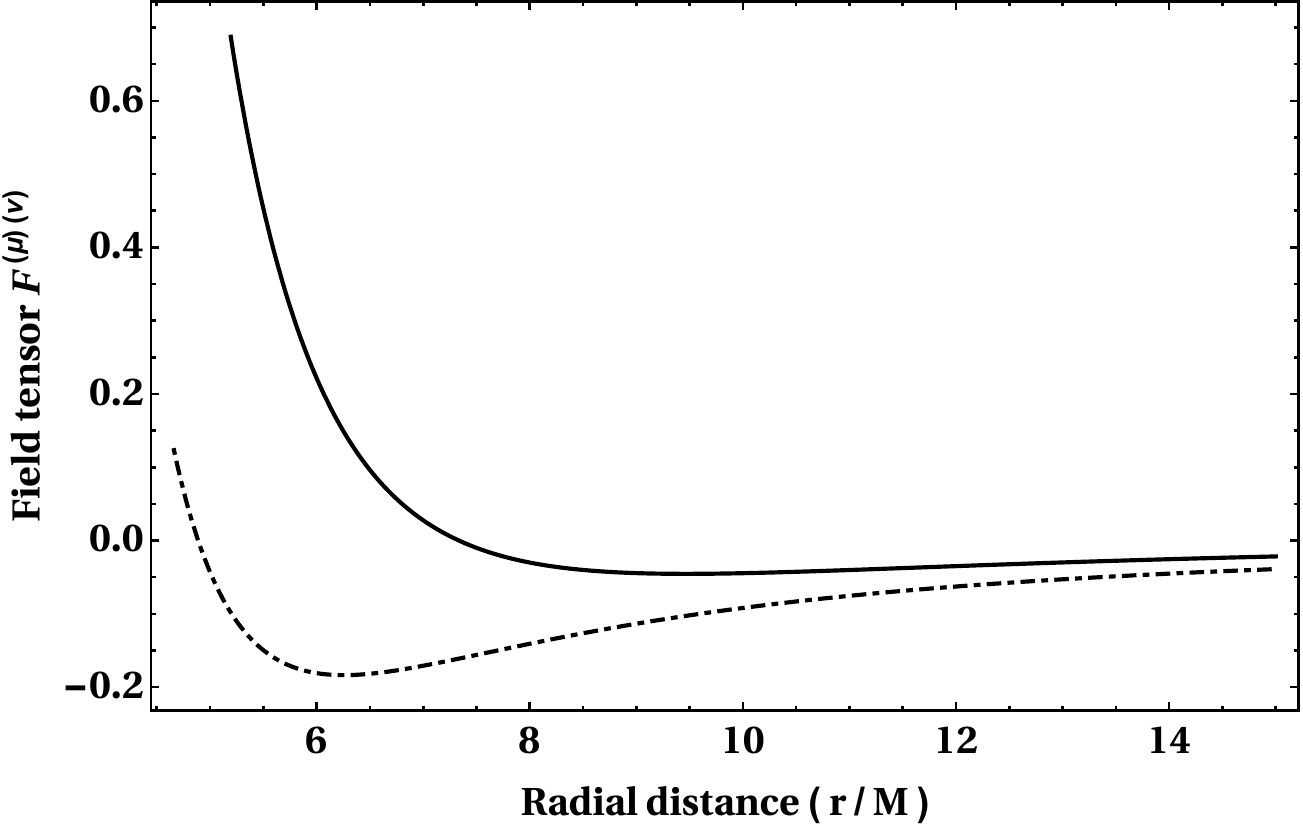}
\caption{In the above figure, the solid and dashed curves describe the $F^{(0)(1)}$, and $F^{(2)(3)}$ components respectively, for $l=3M$, $C_1=2M^3$ and $C_2=4$. Both of these curves consist of extrema and approach zero at the asymptotic infinity.}
\label{fig:field-tensor}
\end{figure}

Motivated by the above discussion, we will follow a similar approach in our case, too. However, before that, we need to introduce the metric ansatz. Given that we will consider a 6-dimensional spacetime ($t,r,\theta_1,\phi_1,\theta_2,\phi_2$) with the non-spherical $S^{(2)} \times S^{(2)}$ topology of the horizon, the metric can be written in the following form:  
\begin{widetext}
\begin{eqnarray}
ds^2=-\dfrac{\Delta}{\rho^2}\Bigl\{dt + P_1 d\phi_1 +P_2d\phi_2 \Bigr\}^2+\dfrac{\rho^2}{\Delta}dr^2+\dfrac{\rho^2}{3} \bigl\{
d\theta_1^2+\sin^2\theta_1 d\phi^2_1+d\theta_2^2+\sin^2\theta_2 d\phi^2_2 \bigr\}, \nonumber \\
\label{eq:metric_S2_S2_SNUT}
\end{eqnarray}
\end{widetext}
where, $P_1=2l\cos\theta_1/3$, $P_2=2l\cos\theta_2/3$, $\rho^2=r^2+l^2$, and $\Delta$ is a function of $r$ only. Note that $\Delta$ can be written further as $\Delta=\rho^2(1-f(r))$, and in the next section, we will employ the field equations to obtain $f(r)$. For any further details on the metric structure for higher-dimensional NUT BHs, we refer our readers to the first part of our study \cite{Mukherjee:2020lld}. Finally, we write the vector potential as 
\begin{equation}
A_{0}=\Psi(r), \quad A_{3}=P_{1}\Psi(r), \quad \text{and} \quad A_{5}=P_{2}\Psi(r).
\end{equation}
For the above choice, only the following covariant components of the field tensor would survive:
\begin{eqnarray}
F_{10}&=&\Psi^{\prime}(r)=\partial_{r} \Psi(r), \quad F_{13} = P_1 F_{10}, \quad F_{15}=P_2 F_{10}, \nonumber \\
F_{23}&=&-\dfrac{2l\sin\theta_1}{3} \Psi(r), \quad F_{45} = -\dfrac{2l\sin\theta_2}{3} \Psi(r). \nonumber \\
\end{eqnarray}
On the other hand, the surviving contra-variant components are given below in matrix form:
\begin{widetext}
\begin{equation}
F^{\mu \nu} = 
\begin{bmatrix}
0 & \Psi^{\prime}(r) & -\dfrac{4 l^2 \cot\theta_1 \Psi(r)}{(r^2+l^2)^2} & 0& -\dfrac{4 l^2 \cot\theta_2 \Psi(r)}{(r^2+l^2)^2}  & 0 \\
-\Psi^{\prime}(r) & 0 & 0 & 0 & 0 & 0 \\
\dfrac{4 l^2 \cot\theta_1 \Psi(r)}{(r^2+l^2)^2}  & 0  & 0 & -\dfrac{6 l \csc\theta_1 \Psi(r)}{(r^2+l^2)^2} & 0 & 0  \\
0 & 0 & \dfrac{6 l \csc\theta_1 \Psi(r)}{(r^2+l^2)^2} & 0 & 0 & 0 \\
\dfrac{4 l^2 \cot\theta_2 \Psi(r)}{(r^2+l^2)^2} & 0 & 0 & 0 & 0 & -\dfrac{6 l \csc\theta_2 \Psi(r)}{(r^2+l^2)^2} \\
0 & 0 & 0 & 0 & \dfrac{6 l \csc\theta_2 \Psi(r)}{(r^2+l^2)^2} & 0 
\label{eq:F_mu_nu}
\end{bmatrix}
\end{equation}
\end{widetext}
We now employ the above components in the field equation as given in \ref{eq:field_eq}. For each of the components of $F^{\mu \nu}$, these equations are given as follows:
\begin{widetext}
\begin{eqnarray}
\partial_{\nu}\left(\sqrt{-g} F^{0 \nu}\right) &= & \dfrac{1}{9} \sin\theta_1 \sin\theta_2 \left \{8 l^2 \Psi(r)+(l^2+r^2)\left[4 r \Psi^{\prime}(r)+(l^2+r^2)\Psi^{\prime \prime}(r) \right]\right \}=0, \label{eq:F0mu} \\
\partial_{\nu}\left(\sqrt{-g} F^{1 \nu}\right)&=&\partial_{\nu}\left(\sqrt{-g} F^{2 \nu}\right)=\partial_{\nu}\left(\sqrt{-g} F^{3 \nu}\right)=\partial_{\nu}\left(\sqrt{-g} F^{4 \nu}\right)=\partial_{\nu}\left(\sqrt{-g} F^{5 \nu}\right)=0.
\end{eqnarray}
\end{widetext}
By solving \ref{eq:F0mu}, we arrive at the following expression for the potential $\Psi(r)$:
\begin{equation}
\Psi(r)=\dfrac{r C_1}{(r^2+l^2)^2}+\dfrac{C_2 \left(r^4+6 l^2 r^2-3 l^4\right)}{3 (l^2+r^2)^2},
\label{eq:pot_Maxwell_Charge}
\end{equation}
where, $C_1$ and $C_2$ are the Maxwell charges. It seems that $C_1$ simply mimics the electric charge. In the vanishing NUT charge limit or $r \rightarrow \infty$, the second term or $C_2$ becomes a constant, and does not affect the metric structure or field tensors. To expose the true nature of these charges, we concentrate on the nonzero components of the field tensor. However, we make use of the orthonormal tetrad field $e^{(\mu)}_{\alpha}$ to simplify these relations. The expressions of the tetrad components are relegated to the Appendix, see Appendix \ref{Appn:tetrad}. The projections of the field tensors on this tetrad basis are defined as $F^{(\mu)(\nu)}= e^{(\mu)}_{\alpha}e^{(\nu)}_{\beta}F^{\alpha \beta}$, and we will have
\begin{widetext}
\begin{eqnarray}
F^{(0)(1)}&=&-\Psi^{\prime}(r)=\dfrac{3C_1l^2+24C_2l^4r-9C_1r^2-8C_2l^2r^3}{3(r^2+l^2)^3},\nonumber \\
F^{(2)(3)}&=&F^{(4)(5)}=\dfrac{2l\Psi(r)}{r^2+l^2}=\dfrac{-2l \Bigl(3C_1 r+C_2 (r^4+6l^2 r^2-3l^4)\Bigr)}{3(r^2+l^2)^3},
\end{eqnarray}
\end{widetext}
while all the other components are zero. These components are demonstrated in \ref{fig:field-tensor}. As shown, both of these curves have an identical nature and consist of an extremum. In the $r \rightarrow \infty$, both of these components vanish. The series expansions of the above expressions, along with the potential, are given as follows:
\begin{widetext}
\begin{eqnarray}
\Psi(r)&=&\dfrac{C_2}{3}+\dfrac{4C_2l^2}{r^2}+\dfrac{C_1}{r^3}-\dfrac{4C_2 l^4}{r^4}-\dfrac{2C_1l^2}{r^5}+\mathcal{O}(1/r^6), \nonumber \\
F^{(0)(1)}&=& -\dfrac{8C_2l^2}{3r^3}-\dfrac{3C_1}{r^4}+\dfrac{16C_2l^4}{r^5}+\dfrac{10C_1 l^2}{r^6}+\mathcal{O}(1/r^7), \nonumber \\
F^{(2)(3)}&=&F^{(4)(5)}= -\dfrac{2C_2l}{3r^2}-\dfrac{2C_2l^3}{r^4}-\dfrac{2C_1l}{r^5}+\dfrac{10C_2l^5}{r^6}+\dfrac{6 C_1 l^3}{r^7}+\mathcal{O}(1/r^8).
\end{eqnarray}
\end{widetext}
The following comments are intended to summarize the properties of these charges:
\begin{itemize}
\item First, $C_2$ behaves as the potential at infinity \cite{Awad:2005ff}, only affects the spacetime geometry in the presence of NUT charge, and takes no part otherwise.

\item In each of these cases, the effects of $C_2$ appears in order lower than $C_1$, which means the $C_2$ has dominant contribution ( in domains $r>M$ ) compare to $C_1$. 

\item The electric field component $F^{(0)(1)}$ is stemmed from both $C_1$ and $C_2$. However, the contribution from $C_1$ appears as $\mathcal{O}(1/r^4)$, while in the potential it comes as $\mathcal{O}(1/r^3)$, which hints that it is a \textit{Coulombic} charge. This also hints that $C_2$ is certainly not a Coulombic charge.

\item Interestingly, in a 4-d spacetime with spherical topology, $\Psi(r)$ can be expanded as follows:
\begin{widetext}
\begin{eqnarray}
\Psi(r)=C_2+\dfrac{C_1-2C_2l}{r}-\dfrac{2C_2l^2}{r^2}+\dfrac{2C_2l^3-C_1l^2}{r^3}+\mathcal{O}(1/r^4), \nonumber \\
\end{eqnarray}
\end{widetext}
which hints that together $C_2$ and $l$ induce a magnetic charge and also affects the Coulombic charge. This feature is absent in higher dimensions, and the electric charge is entirely given by $C_1$, where $C_2$ takes no part in shaping it. 

\item Besides, it should be recalled that given the potential is dimensionless, the dimension of $C_1$ is $\text{length}^3$ and $C_2$ is simply dimensionless.
\end{itemize}
It is now time that we should introduce the stress-energy tensor of the electromagnetic field explicitly (given in \ref{eq:Energy_momentum}), and below we present them term by term:
\begin{eqnarray}
{T}^{(0)}_{(0)} &=& 2 \left[F^{(0) (i)}F_{(0) (i)}\right]-(1/2)F^{(\mu) (\nu)}F_{(\mu) (\nu)}, \nonumber\\
{T}^{(1)}_{(1)} &=& 2 \left[F^{(1) (\mu)}F_{(1) (\mu)}\right]-(1/2)F^{(\mu) (\nu)}F_{(\mu) (\nu)},\nonumber\\
{T}^{(i)}_{(i)} &=& 2 \left[F^{(i) (\mu)}F_{(i) (\mu)}\right]-(1/2)F^{(\mu) (\nu)}F_{(\mu) (\nu)},
\nonumber
\end{eqnarray}
where $i$ only runs in the spatial indices and \enquote*{bracket} denotes a quantity projected on the tetrad frame. As the dimension is six, the trace of the stress-energy tensor is nonzero, and given as
\begin{equation}
{T}={T}^{(0)}_{(0)}+{T}^{(1)}_{(1)}+\sum_{i=2}^{4}{T}^{(i)}_{(i)}.
\end{equation}
From the condition of static spacetime, we further have ${T}^{(0)}_{(0)}={T}^{(1)}_{(1)}$, and ${T}^{(2)}_{(2)}={T}^{(3)}_{(3)}={T}^{(4)}_{(4)}={T}^{(5)}_{(5)}$. Remember, in the 4-dimensional sapcetime, we also have ${T}^{(1)}_{(1)}=-{T}^{(2)}_{(2)}$ and therefore, the trace would be zero. However, in the present case, ${T}^{(1)}_{(1)} \neq -{T}^{(2)}_{(2)}$, and trace is given by
\begin{equation}
T=2{T}^{(1)}_{(1)}+4 {T}^{(2)}_{(2)}.
\end{equation}
Finally, by the use of Maxwell's field tensor, we obtain
\begin{widetext}
\begin{eqnarray}
T^{(1)}_{(1)}&=&T^{(0)}_{(0)}=\dfrac{1}{9(r^2+l^2)^6}\Bigl\{192 C_1 C_2 l^2 r^3 (l^2-r^2)-9 C_1^2(l^4+2l^2 r^2+9r^4)- \nonumber \\
& & \hspace{7cm} 8 C_2^2l^2 \bigl(9l^8+36 l^6 r^2-18 l^4 r^4+20 l^2r^6+r^8\bigr)\Bigr\}, \label{eq:T11_EM}\nonumber \\
T^{(i)}_{(i)}&=&\dfrac{1}{9(r^2+l^2)^6}\bigl(3 C_1 l^2+24 C_2 l^4 r-9 C_1 r^2-8 C_2 l^2 r^3\bigr)^2, \nonumber \\
T &=& \dfrac{2}{9(r^2+l^2)^6}\Bigl\{9C_1^2(9r^4-14l^2 r^2+l^4)-8 C_2^2 l^2(9l^8-108^6 r^2+78l^4 r^4+4l^2 r^6+r^8) \nonumber \\
&& \hspace{8cm} +96C_1 C_2 l^2 r(3l^4-8l^2 r^2+r^4)\Bigr\}. 
\label{eq:Trace_EM}
\end{eqnarray}
\end{widetext}
The above relations are employed to obtain $f(r)$, which we carry out in the next section.
\section{Black hole solutions with NUT and Maxwell charges} \label{sec:obtain_metric}
In the presence of Maxwell's fields, the field equations projected on the orthonormal basis become
\begin{equation}
H^{(a)}_{~(b)}=\alpha_2 \bigl(2 J^{(a)}_{~(b)}-\dfrac{1}{2}\delta^{(a)}_{~(b)}\mathcal{L}_{\rm GB} \bigr)=-\Lambda \delta^{(a)}_{~(b)}+T^{(a)}_{~(b)}.
\end{equation}
The explicit expression for $\mathcal{L}_{ \rm GB}$ and $J^{ab}$ are given by \ref{eq:LGB} and \ref{eq:Jab} respectively. We now substitute the metric ansatz as given in \ref{eq:metric_S2_S2_SNUT} and obtain the expressions for the Gauss-Bonnet tensors and Lagrangian as follows:
\begin{widetext}
\begin{eqnarray}
H^{(0)}_{(0)}&=&  H^{(1)}_{(1)}= \dfrac{12 \alpha_2}{(r^2+l^2)^3}\Bigl\{2 f(r) \Bigl[2 l^2+r (l^2-r^2)f'(r) \Bigr]-2 \Bigl[ r^2+3 l^2+2 l^2 r f'(r) \Bigr] \label{eq:H00}-(r^2+l^2) [f(r)]^2 \Bigr \}, \nonumber \\
H^{(i)}_{(i)} &=& \dfrac{6 \alpha_2}{(r^2+l^2)^3} \Bigl \{4 l^2 r f'(r) +(l^4-r^4) \Bigl[f'(r) \Bigr]^2 -2 l^2 \Bigl[(r^2+l^2)f'(r)-2 \Bigr]\nonumber\\
~&& ~~~~~~~~~~~~~~~~~~~~~~~~~~~+~
f(r) \Bigl[(l^4-r^4)f''(r)-2r (3 l^2 +r^2)f'(r)-4 l^2 \Bigr] \Bigr\},  \nonumber \\
L_{\rm GB} &=& \dfrac{24}{(r^2+l^2)^2}\Bigl\{2+f(r)^2+(r^2-l^2)f'(r)^2+2 l^2 f''(r)+f(r)\Bigl[4 r f'(r)+(r^2-l^2)f''(r)\Bigr]\Bigr\}, \label{eq:LGB_NUT} 
\end{eqnarray}
where, $i$ runs from $2$ to $5$. We would like to mention that we have used $\Delta=\rho^2(1-f(r))$, as also mentioned before, in the line-element of \ref{eq:metric_S2_S2_SNUT} to obtain these expressions. Moreover, from \ref{eq:LGB} and \ref{eq:Jab} we observe that $\mathcal{L}_{ \rm GB}$ and $J^{ab}$ are obtained from the Riemann and Ricci tensors and Ricci scalar, which are again obtained by taking derivative of the metric tensor. Thus, in the final expression of \ref{eq:LGB_NUT}, we have derivatives of $f(r)$. Given that the expression of $H^{(0)}_{(0)}$ contains first-order derivative of $f(r)$, we can solve the following
\begin{equation}
H^{(0)}_{(0)}-T^{(0)}_{(0)}+\Lambda=0.
\label{eq:field_EM}
\end{equation}
By plugging in the expressions of $T^{(0)}_{(0)}$ and $H^{(0)}_{(0)}$ from  \ref{eq:Trace_EM} and \ref{eq:H00} respectively, we can solve \ref{eq:field_EM}, and obtain $f(r)$ 
\begin{eqnarray}
f(r) &=& \dfrac{1}{90(l^4-r^4)}\Bigl[180 l^2 (r^2+l^2)-\sqrt{\dfrac{15}{\alpha_2}}\Bigl\{480 C_1 C_2 l^2 r^3 (l^2-r^2)+45 C_1^2(l^4+2 l^2 r^2-3 r^4) \nonumber \\
& & +40 C_2^2 l^2 \left(9 l^8+6 l^4 r^4-16 l^2 r^6+r^8\right)+9 (r^2+l^2)^2\Bigl[5 l^8 \Lambda-20 l^6 r^2 \Lambda +10 l^4 (-12 \alpha_2 +r^4 \Lambda)\nonumber \\
& & + r^3 \bigl(60 M \alpha_2-120 r \alpha_2+r^5 \Lambda \bigr)+4 l^2 r \bigl(-15 M \alpha_2+120 r \alpha_2+r^5 \Lambda\bigr) \Bigr] \Bigr\}^{1/2}\Bigr].
\label{eq:f(r)_charged}
\end{eqnarray}
\end{widetext}
where $M$ is the integration constant. Without losing any generality, we set the value of $\alpha_2=M^2$, to keep it consistent with its dimension. For a clear exposition of our results, we rewrite \ref{eq:f(r)_charged} as follows:
\begin{equation}
f(r)=\dfrac{180l^2(r^2+l^2)-\left\{15 h(r)/M^2 \right\}^{1/2}}{90(l^4-r^4)},
\label{eq:f_r_h_r}
\end{equation}
The term inside the root is given as $h(r)$,
\begin{widetext}
\begin{eqnarray}
h(r)&=&480 C_1 C_2 l^2 r^3 (l^2-r^2)+45 C_1^2(l^4+2 l^2 r^2-3 r^4) \nonumber \\
& & +40 C_2^2 l^2 \left(9 l^8+6 l^4 r^4-16 l^2 r^6+r^8\right)+9 (r^2+l^2)^2\Bigl[5 l^8 \Lambda-20 l^6 r^2 \Lambda +10 l^4 (-12M^2 +r^4 \Lambda)\nonumber \\
& & \hspace{3.9cm} + r^3 \bigl(60 M^3-120 r M^2+r^5 \Lambda \bigr)+4 l^2 r \bigl(-15 M^3+120 r M^2+r^5 \Lambda\bigr) \Bigr].
\label{eq:hr_general_first}
\end{eqnarray}
\end{widetext}
To have a real solution, we must have $h(r)\geq 0$. Using the above, we can now attempt to study some of the limiting cases of the present solution and realize whether our solution matches with the existing literature or not. 
\begin{enumerate}
\item In the limit of vanishing NUT charge, i.e., ${l}=0$, we retrieve the following expression:
\begin{equation}
f({r})\bigg|_{{l}=0}=\sqrt{\dfrac{M}{{r}}-2+\dfrac{{\Lambda} {r}^4}{60 M^2}-\dfrac{{C_1}^2}{4{r}^4 M^2}},
\end{equation}
which is independent of the Maxwell's charge $C_2$. This is expected as in the $l=0$ limit, the second term in the potential (see \ref{eq:pot_Maxwell_Charge}) behaves as a gauge and makes no contribution to the stress-energy tensor or field equations. On the contrary, $C_1$ behaves as the electric charge and comes with a $\sim 1/r^4$ contribution in 6-d spacetime. Besides, it is important to note that in the asymptotic limit, i.e., $ r \rightarrow \infty$, the above solution is only valid in the presence of a cosmological constant. In the $C_1=0$ case, the above expression matches with Eq. (3.2) in Ref. \cite{Dadhich:2015nua}. 

\item  In the case of ${r}$ approaches infinity, $f({r})$ becomes
\begin{eqnarray}
f({r})\bigg|_{{r} \rightarrow \infty} = \Bigl\{\dfrac{\Lambda}{60M^2} \Bigl(19l^4+6l^2r^2+r^4\Bigr)+\nonumber\\
\dfrac{2 C_2^2 l^2}{27M^2}-2\Bigr\}^{1/2},
\label{eq:f(r)_S2XS2_r_infinity}
\end{eqnarray}
and it indicates the qualitative difference between $C_1$ and $C_2$. For a nonzero $C_2$, it is not essential to have a nonzero $\Lambda$ to consistently describe the asymptotic limit, as far as $C_2^2 l^2> 27 M^2$ is satisfied. This is in stark contrast to $C_1$, which does not change the asymptotic limit of our solution. 

\item With a simple substitution of $r=l$, it may seem that the above equation would diverge. However, if we consider the limit properly, we arrive at the following expression:
\begin{eqnarray}
f({r})\bigg|_{{r} = l} &=& 1+\dfrac{M}{4l}+\dfrac{\Lambda l^4 }{15 M^2}-\dfrac{C_1^2}{48 l^4 M^2}-\nonumber \\
&& \hspace{2.5cm} \dfrac{4 C_2^2 l^2}{27 M^2}-\dfrac{C_1 C_2}{18 l M^2},
\label{eq:f(r)_r=l}
\end{eqnarray}
which is regular as far as $l \neq 0$. 
\end{enumerate}
\subsection{Non-central singularity}
The non-spherical product horizon topology in higher dimensions introduces a non-central singularity where the Ricci and Kretschmann scalars diverge. In part-I of our study \cite{Mukherjee:2021erg}, we discuss this feature in pure \GB~NUT solutions, following the analysis given in Ref. \cite{Pons:2014oya}. This singularity is unphysical in nature, and has no proper reason for occurrence. Therefore, it needs to be avoided to arrive at a consistent BH solution. One way is to hide this singularity behind the event horizon, or choose a parameter space such that it does not appear in the first place. It is expected that, in either of these cases, the BH parameters may be severely conditioned. The location of this singularity is given by the real roots of the equation $h(r)=0$. Therefore, to avoid singularity and describe a BH solution, we must have $h(r)>0$.
\subsection{Event \& cosmological horizons} \label{sec:Ev_Cos_Hor}
The locations of the horizon are given by the real roots of the solution $f(r)=1$, which is obtained from $\Delta=\rho^2(1-f(r))=0$. This can also be written in terms of $h(r)$ as follows:
\begin{equation}
 h(r)=540M^2(r^2+l^2)^4.
\end{equation}
Note that if we substitute the above in \ref{eq:f_r_h_r}, both the numerator and denominator vanish at the $r=l$ limit. Therefore, the above equation is always trivially satisfied for $r=l$, which, however, may not correspond to a horizon solution. This can also be understood by referring to \ref{eq:f(r)_r=l}, which trivially does not produce a horizon unless we have 
\begin{equation}
 \dfrac{M}{4l}+\dfrac{\Lambda l^4}{15M^2}-\dfrac{C^2_1}{48 M^2 l^4}-\dfrac{C_1 C_2}{18 l M^2}-\dfrac{4 C^2_2l^2}{27M^2}=0. 
\end{equation}
Interestingly, the above can never be satisfied without the presence of Maxwell charges as $\Lambda$ is always positive. In the next section, we will discuss the solutions of the above equation for different Maxwell and NUT charges.
\section{Validity of the solution} \label{sec:Validity}
%
%
In order to have a organized study, we consider the following cases, namely, (A) when $C_1 \neq 0$, $C_2=0$, (B) when $C_1=0$, $C_2 \neq 0$, and (C) when both the charges are non-zero, $C_1 \neq 0$, $C_2 \neq 0$ which constitutes a general study. Finally, in (D), we will assume a case with $\Lambda=0$.
\subsection{For $\mathbf{C_1 \neq 0, C_2=0}$}
\begin{figure*}
\centering
\includegraphics[scale=0.35]{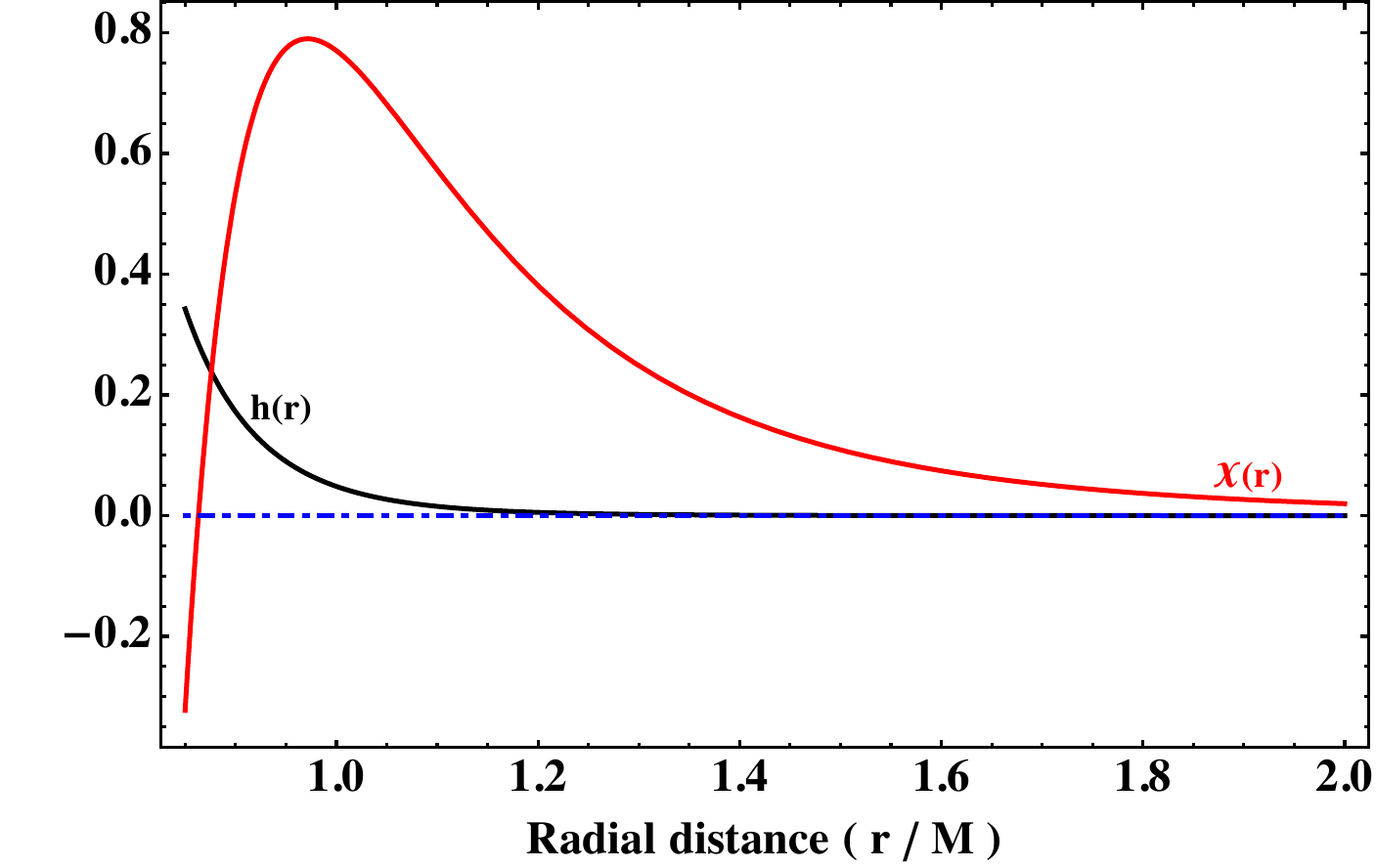}
\hskip 10pt
\includegraphics[scale=0.35]{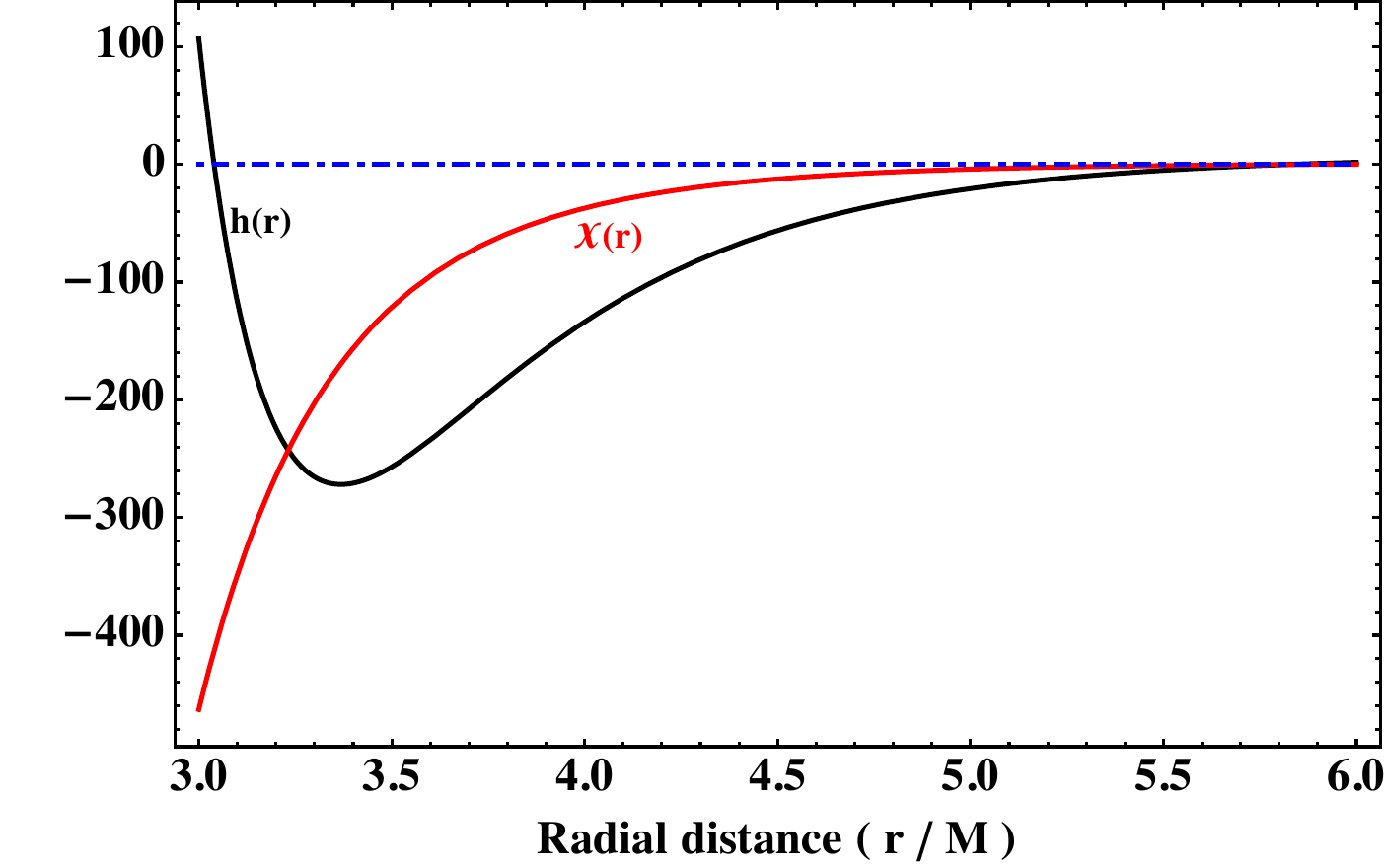}
\caption{Above demonstrates the spacetime structure for $l=3M$, $\lambda=30$, $C_2=0$. The left figure is for $C_1=30M^3$, and have one horizon (event) located at $r_{\rm h}=0.86M$ and no spacetime singularity. In the right plot, we set $C_1=400M^3$, and the spacetime contains two singularities at $r_{\rm c}=3.04M, 5.83M$, and the singularities are covered by the event horizon at $r_{\rm h}=5.93M$.}
\label{Figure_case_2}
\end{figure*}

To investigate the effects of Maxwell's charges, we consider the case with $C_2=0$ first, and concentrate on the Coulombic counterpart, i.e, $C_1 \neq 0$. Note that we need to find a window of parameters for which $h(r)>0$ is always satisfied outside the event horizon. With $C_2=0$, the expression for $h(r)$ becomes:
\begin{widetext}
\begin{eqnarray}
h(r) &=& 9 \Lambda r^{12}+54 \Lambda l^2 r^{10}+9 (19 l^4 \Lambda-120 M^2)r^8+540 M^3 r^7+36l^2(l^4 \Lambda+60M^2)r^6+540 M^3 l^2 r^5 \nonumber \\
& & -45 (3 C_1^2+5 l^8 \Lambda-144 M^2 l^4)r^4-540 M^3 l^4 r^3+90 l^2 (C_1^2-l^8 \Lambda+24 l^4 M^2)r^2-540 M^3 l^6 r \nonumber \\
& &  \hspace{10cm}+ 45 l^4 (C_1^2+l^8 \Lambda-24 l^4 M^2). \nonumber \\ 
\label{eq:hr_Clmb}
\end{eqnarray}
The locations of the horizon, $r_{\rm h}$, given by the equation $h(r_{\rm h})=540 M^2(r_{\rm h}^2+l^2)^4$ and it reads as
\begin{equation}
 h(r)-540M^2(r^2+l^2)^4=9(r^2-l^2)\chi(r).
\end{equation}
Based on the discussions in \ref{sec:Ev_Cos_Hor}, we conclude that the solution of the above equation is given by
\begin{eqnarray}
\chi(r)&=& \Bigl\{\Lambda r^{10}+7 l^2 \Lambda r^8+(26 \Lambda l^4-180 M^2)r^6+60 M^3 r^5+30 l^2(\Lambda l^4-6M^2)r^4 \nonumber \\
& & \hspace{4.5cm} +120 M^3 l^2 r^3-5 (3 C_1^2-\Lambda l^8-36 l^4 M^2)r^2+60 M^3 l^4 r \nonumber \\
& & \hspace{8cm}-5 l^2(C_1^2+\Lambda l^8-36 l^4M^2)\Bigr\}=0. \nonumber \\ 
\label{eq:chir}
\end{eqnarray}
\end{widetext}
With these expressions, we now attempt to obtain the range of viable parameters for the solution. For an illustration, we assume $\lambda= \Lambda l^4/M^2$, and display a couple of cases below in \ref{Figure_case_2}.
\subsection{For $\mathbf{C_1 = 0, C_2 \neq 0}$}
\begin{figure*}
\centering
\includegraphics[scale=0.35]{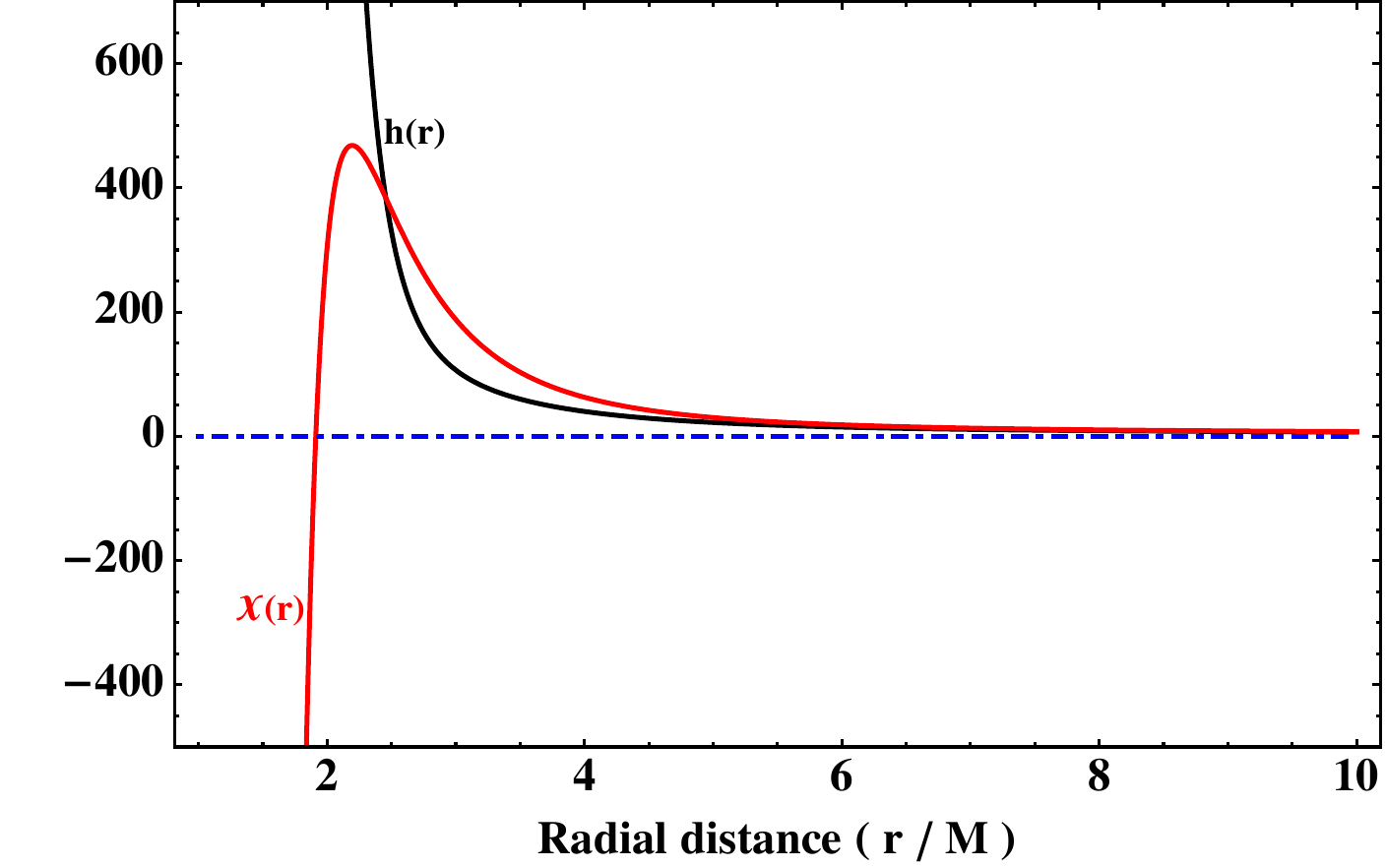}
\hskip 10pt
\includegraphics[scale=0.42]{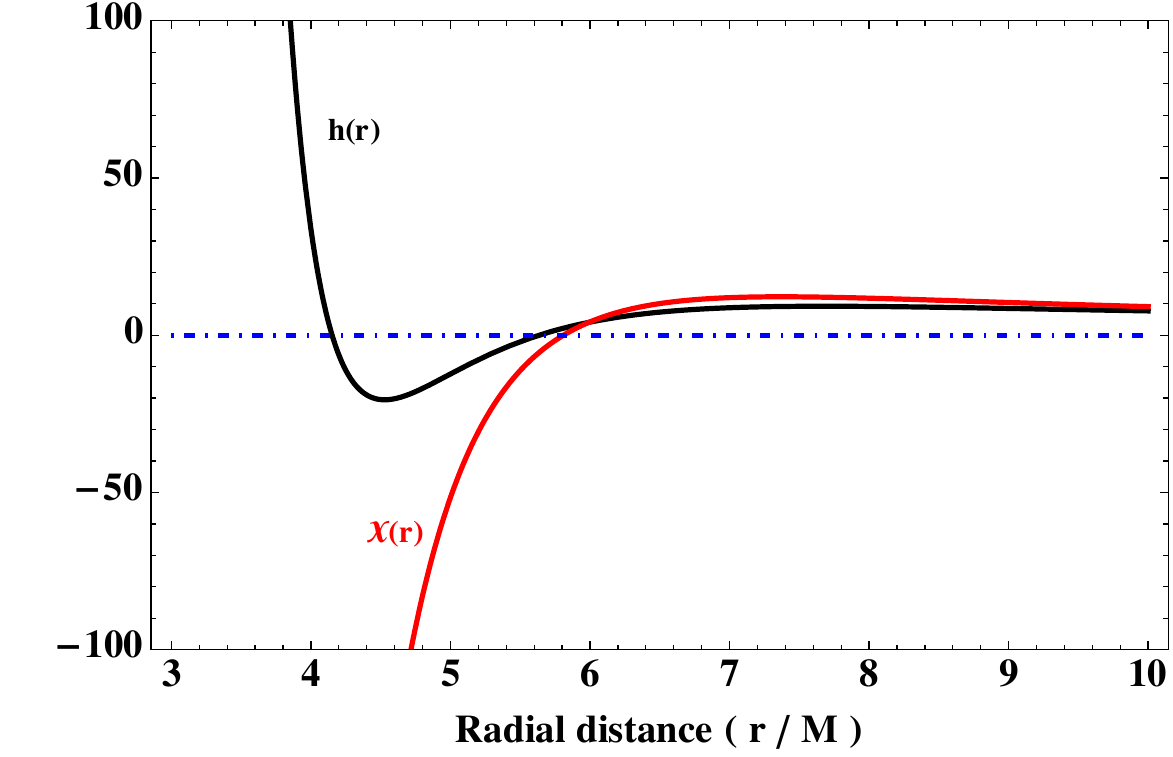}
\caption{In top left: Here we have chosen $l=3M$, $\lambda=38$ and $C_2=0.8$, and depict scaled values of $h(r)$ and $\chi(r)$. There is no singularity, and the event horizon is at $r_{\rm h}=1.91185M$. Top right: The scaled values of $h(r)$ and $\chi(r)$ are shown for $l=4M$, $\lambda=100$ and $C_2=3$. In this case, the singularities are $4.1519 M$ and $5.6345 M$, which are covered by the horizon at $5.8094M$.}
\label{fig:Figure_case_20} 
\end{figure*}

The presence of $C_2$ and NUT charge imparts a nontrivial contribution to shaping the vector potential. In other words, while $C_2$ behaves as a gauge and has no effect on the metric components whenever $l=0$, the same is not true for $l \neq 0$. In order to stress these effects of $C_2$, we set $C_1$ to zero and study the spacetime structure. With this, we arrive at the following expression for $h(r)$
\begin{widetext}
\begin{eqnarray}
h(r) &=& 9 \Lambda r^{12}+54 \Lambda l^2 r^{10}+\bigl(171 l^4 \Lambda+40 l^2 C_{2}^2-1080 M^2 \bigr)r^8+540 M^3 r^7+4 l^2 \bigl(540M^2-160 C_2^2 l^2 \nonumber\\
& & +9 \Lambda l^4 \bigr)r^6+ 540 M^3 l^2 r^5+15 l^4 r^4\bigl(432M^2+16 C_2^2 l^2-15 \Lambda l^4 \bigr)-540M^3  l^4 r^3-90 l^6r^2 (\Lambda l^4-24 M^2)-\nonumber \\ 
& & \hspace{9cm} 540 M^3 l^6 r+45 l^8 (8 C_2^2 l^2+\Lambda l^4-24M^2).\nonumber 
\\
\label{eq:C_1_0}
\end{eqnarray}
\end{widetext}
The other important equation is the relation for the location of the horizon, and it is given as
\begin{widetext}
\begin{eqnarray}
h(r)-540M^2 (r^2+l^2)^4=(r^2-l^2)\Bigl\{9 \Lambda r^{10}+63 l^2 \Lambda r^8+(234 l^4 \Lambda+40 C_2^2 l^2-1620 M^2)r^6+540 M^3 r^5 \nonumber \\
-30 r^4 (54 l^2 M^2+20 C_2^2 l^4-9 \Lambda l^6)+1080 M^3 l^2 r^3+45 l^4 r^2(36M^2-8 C_2^2l^2+l^4 \Lambda)+\nonumber \\
540M^3 l^4 r-45 l^6 (l^4 \Lambda+8 C_2^2 l^2-36M^2)\Bigr\}. \nonumber \\
\label{eq:horizon_C1_0}
\end{eqnarray}
\end{widetext} 
We should start by recalling that the term outside the bracket is not useful as far as we are concerned to obtain the locations of the horizon. In \ref{fig:Figure_case_20}, we have plotted two cases where the location of the horizon is shown.
%
%
%
%
%
\subsection{For $\mathbf{C_1 \neq 0, C_2 \neq 0}$} \label{sec:c1neqc2neq}
\begin{figure*}
	\centering
	\includegraphics[scale=0.35]{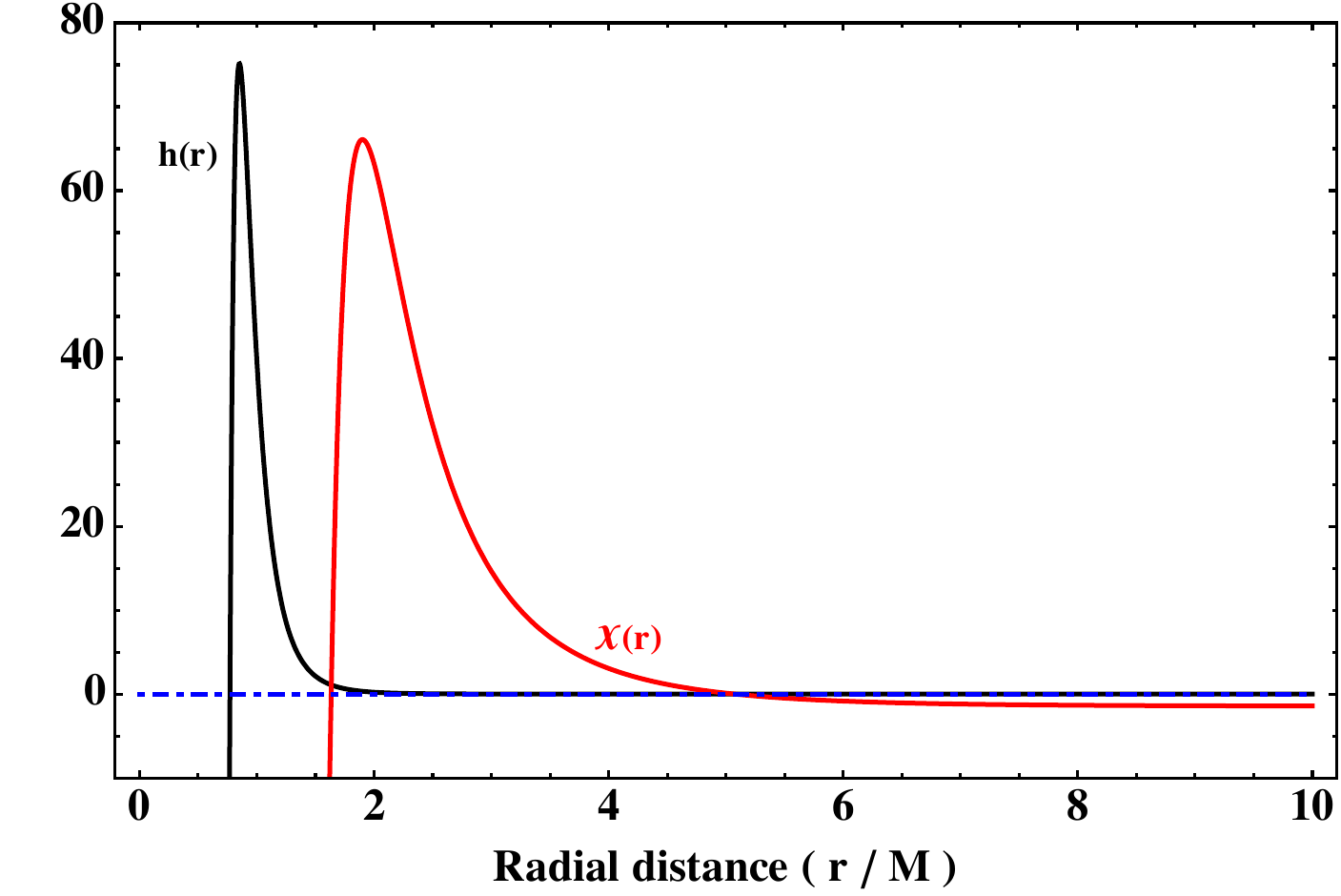}
    \hskip 10pt
	\includegraphics[scale=0.35]{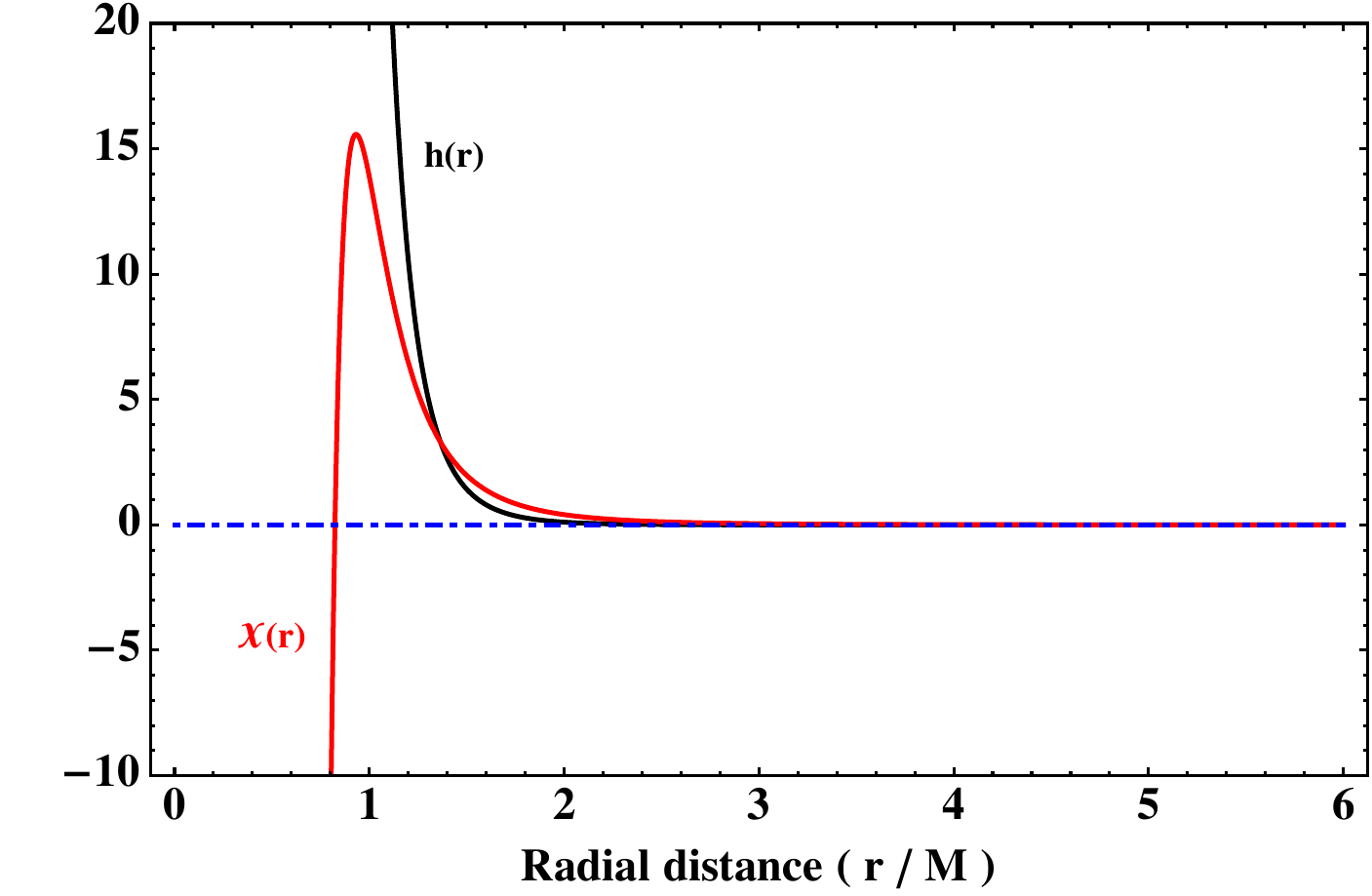}
	\caption{The above figures demonstrate the spacetime structure for $l=1.5M$ and $\Lambda M^4=0.15$, and both are physically viable solutions. On the left, we set $C_1=M^3$ and $C_2=0.5$, and the singularities are covered by the event horizon. On the right, we have $C_1=10M^3$ and $C_2=0.5$, and in this case, there exists no singularity, but two horizons.}
\label{Figure_16R}
\end{figure*}

When both of the Maxwell charges are set to nonzero, it may be considered as the most general case compared to the other scenarios we studied till now. In this case, the expression for $h(r)$ is given by \ref{eq:hr_general_first}. However, for our purpose, we have rewritten that expression by expanding it in the powers of $r$ as follows
\begin{widetext}
\begin{eqnarray}
h(r)=9 \Lambda r^{12}+54 \Lambda l^2 r^{10}+(171 l^4 \Lambda+40 C_2^2l^2-1080M^2)r^8+540M^3r^7+4l^2(9\Lambda l^4-160C_2^2 l^2+540M^2)r^6 \nonumber \\
+60 l^2 r^5(9M^3-8C_1C_2)-15 (9C_1^2-432l^4M^2-16 C_2^2 l^6+15 l^8 \Lambda)r^4+60 l^4 r^3(-9M^3+8C_1C_2)+\nonumber \\
90 l^2 r^2(C_1^2-l^8 \Lambda+24 l^4M^2)-540M^3 l^6 r+45 l^4(C_1^2-24 l^4M^2+8C_2^2l^6+l^8 \Lambda), \nonumber \\
\label{eq:C_2_C_1}
\end{eqnarray} 
\end{widetext}
which reduces to the known equations in relevant limits. The other equation that gives the location of horizons, is given as follows:
\begin{widetext}
\begin{eqnarray}
h(r)-540 M^2(r^2+l^2)^4=(r^2-l^2) \Bigl\{9 \Lambda r^{10}+63 \Lambda l^2 r^{8}+(40 C_2^2l^2+234 l^4 \Lambda-1620M^2)r^6+540M^3 r^5 \nonumber \\
-30 (54 l^2M^2+20 C_2^2 l^4-9l^6 \Lambda)r^4+120l^2(9M^3-4C_1C_2)r^3-45(3C_1^2-36l^4M^2+8C_2^2l^6-l^8 \Lambda)r^2+\nonumber \\
540M^3l^4 r-45 l^2(C_1^2-36l^4M^2+8 C_2^2l^6+l^8 \Lambda)\Bigr\}.
\end{eqnarray}
\end{widetext}
In either of the above cases, it is hard to argue about the coefficients of various orders due to the larger set of parameters. However, the underlying machinery remains the same as we already discussed in earlier cases. Therefore, we simply highlight two cases and demonstrate them in \ref{Figure_16R} for $l=1.5M$ and $\Lambda M^4=0.15$. Different values of $C_1$ and $C_2$ represent different curves and contain different information. For example, in the left, we set $C_1=M^3$ and $C_2=0.5$, and the spacetime singularity exists at $r_{\rm c}=0.771424M$. This singularity is covered with event horizon at $r_{\rm h}=1.6345M$, while the other horizon is located at $5.07105M$. On the right, we have $C_1=10M^3$ and $C_2=0.5$, whereas the spacetime is free of singularity and consists of both event and cosmological horizons.
\subsection{For vanishing cosmological constant $\mathbf{\Lambda=0}$}
\begin{figure*}
	\centering
	\includegraphics[scale=0.4]{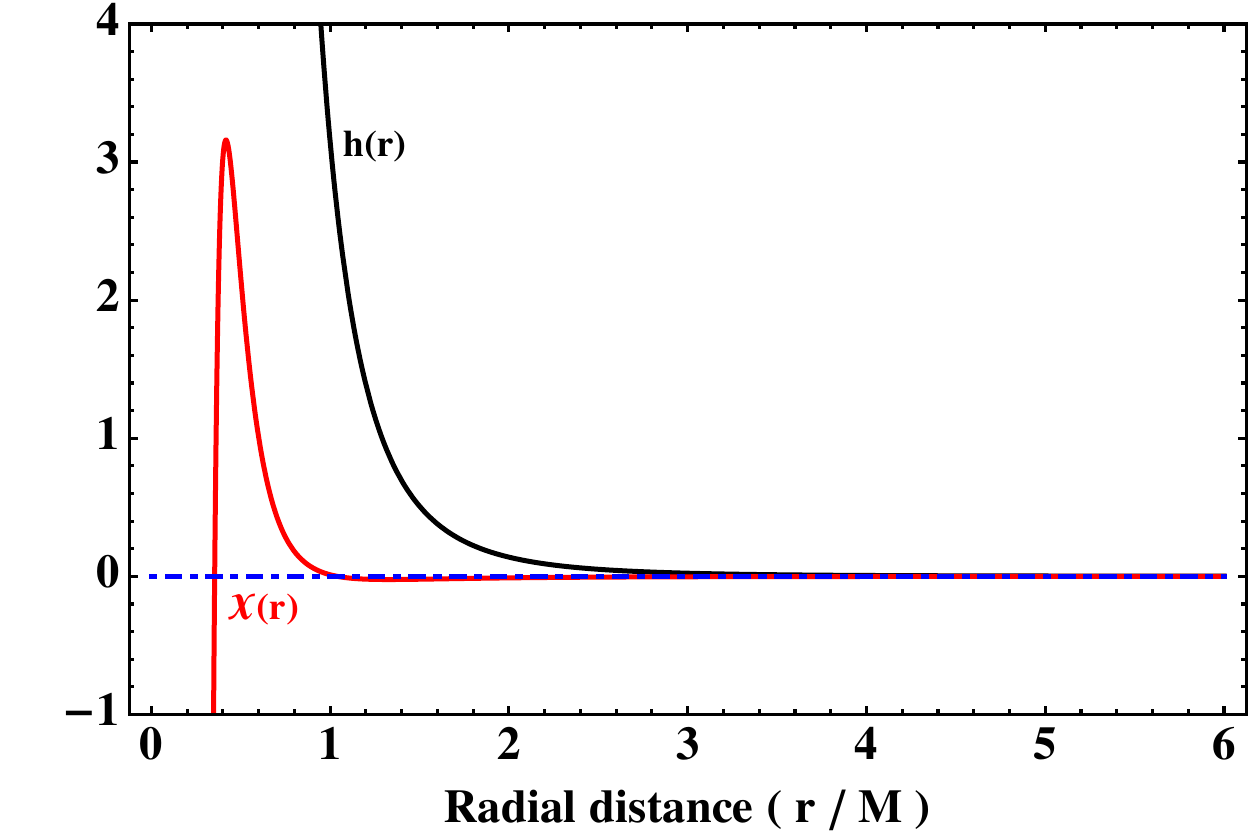}
    \hskip 10pt
	\includegraphics[scale=0.4]{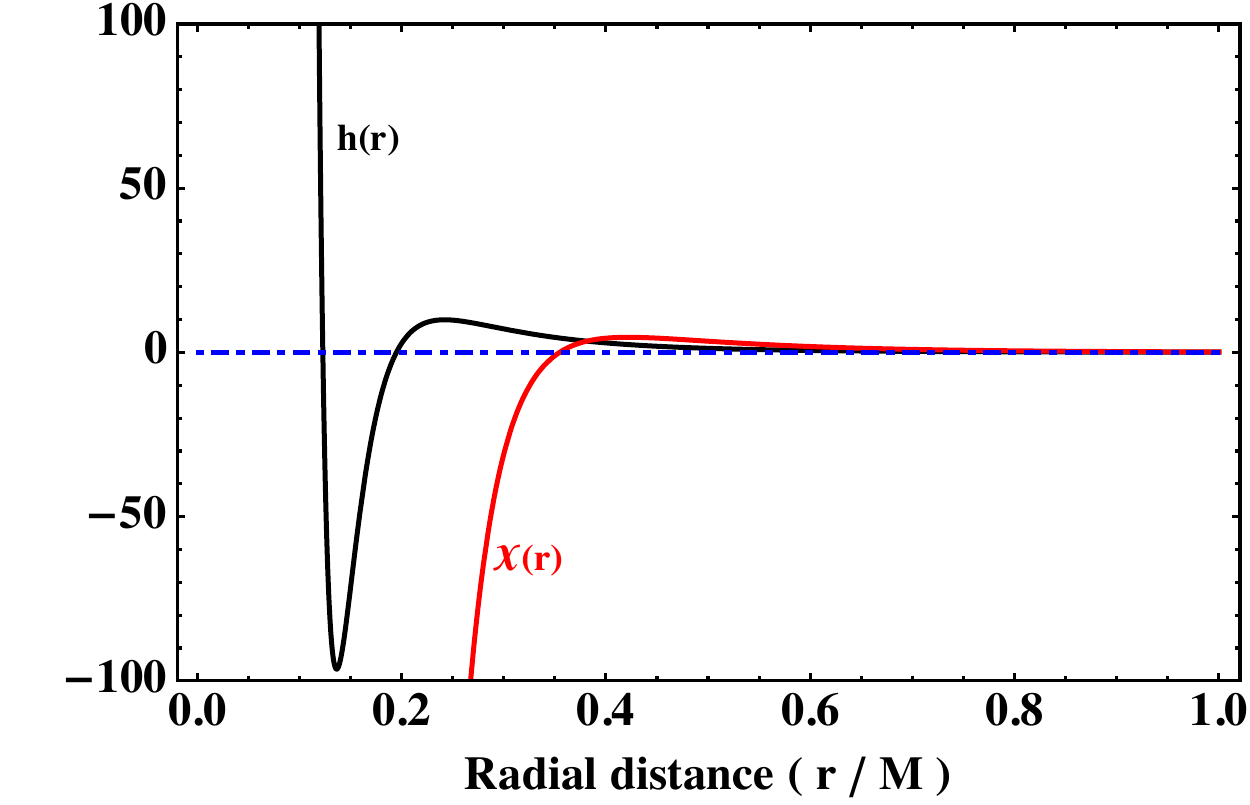}
	\caption{The above figures demonstrate the spacetime structure for $l=0.09M$, $C_1=0$ and $\Lambda M^4=0$, for different values of $C_2$. In the top right, we have $C_2=62$, and the spacetime is viable and free of singularity. The inner and outer horizons are located at $0.354242M$ and $1.03962M$ respectively. For the top right, we set $C_2=65$, and the singularities reappear. However, the singularities located at $0.122875M$ and $0.1949$, are always covered by the inner and outer horizons located at $0.354635M$ and $1.75391M$ respectively.}
\label{Figure_17R}
\end{figure*}

\noindent
Referring to \ref{eq:f(r)_S2XS2_r_infinity}, it is easy to realized that $\Lambda$ is not essential when it comes to describe the spacetime in the presence of $C_2$. Of course, there would be a stringent constraint on $C_2$, allowing only the option $C_2^2l^2>27M^2$. In order to outline this phenomena elaborately, we set $C_1=0$ (along with $\Lambda=0$) and arrive at the following expression for $h(r)$:
\begin{widetext}
\begin{eqnarray}
h(r) &= & 20 \Bigl[ 2(C_2 ^2 l^2-27M^2)r^8+27 C_0 r^7+4 l^2(27M^2-8 C_2^2 l^2)r^6+27 M^3 l^2 r^5\nonumber \\
& & +12 l^4(27M^2+C_2^2 l^2)r^4-27 M^3 l^4 r^3+108 l^6 r^2M^2-27 M^3 l^6 r+18 l^8 (C_2^2 l^2-3M^2) \Bigr].
\label{eq:lambda=0}
\end{eqnarray} 
\end{widetext}
A quick look would reveal that the above equation is accompanied with $6$ changes of sign as far as $C_2^2l^2>27M^2$ is satisfied. This would ensure that there can be at most $6$ or less than an even number of real positive solutions, which also includes $0$. The other crucial expression relates the locations of the horizon, and can be written in the following form
\begin{widetext}
\begin{eqnarray}
h(r)-(r^2+l^2)^4=20(r^2-l^2) \Bigl\{(2 C_2^2 l^2-81M^2)r^6+27M^3 r^5-3 l^2(27M^2+10C_2^2 l^2)r^4+54 M^3 l^2 r^3+ \nonumber \\
9 l^4(9M^2-2 C_2^2l^2)r^2+27M^3l^4 r+9 l^6(9M^2-2C_2^2 l^2)\Bigr\}. \nonumber \\
\end{eqnarray} 
\end{widetext}
It is easy to figure out that with the condition $C_2^2 l^2>27 M^2$, the coefficients of $r^2$ and $r^0$ introduce an even change of signs. The only uncertainty is introduced by the coefficient of $r^6$, i.e., whether $2C_2^2l^2>81M^2$ or $2C_2^2l^2<81M^2$. In the first case, there can be a total of $5$ changes of sign, which ensures that the above equation is bound to have one real positive solution. For the latter case, there are an even number of changes attributed to an even number of real and positive solutions. In \ref{Figure_17R}, we illustrate an example where the above-mentioned features become explicit. We assume $l=0.09M$, which further claims $C_2 > 57.735$ from the asymptotic condition $C_2^2l^2>27M^2$. The other details of the plots are given in the caption of the figure. 
%
\noindent
With this, we finish our discussion to inform about various possibilities of black hole solutions and study the parameter space with their prospective physical bounds. Certainly, the interplay between NUT and Maxwell charges introduces many nontrivial properties. We attempt to illustrate the possibilities where the spacetime contains a horizon structure, and either free of singularity or they are hidden behind the horizon. 
\section{Thermodynamic properties} \label{sec:thermo}

In this section, we investigate a few crucial properties of this black hole spacetime that, as we will see, will help us study its horizon thermodynamics. In this regard, we shall first discuss the velocity of a zero angular momentum observer (ZAMO) in this spacetime, which is essential in estimating the horizon temperature. We will also evaluate the expression for the temperature corresponding to a Hawking quanta emitted from the horizon.
\subsection{Zero angular momentum observer (ZAMO)}
Before delving into the thermodynamic aspects, let us revisit a well-known relativistic phenomenon, namely ZAMO. The ZAMO represents an observer with zero angular momentum \cite{schutz2009first}. In the case of a static spacetime such as Schwarzschild, a ZAMO surface also has vanishing angular velocity. However, for a stationary spacetime like Kerr, the same is not true, and we can have nonzero angular velocity. In the present case, the metric consists of cross terms between $t$ and $
\phi$, which will introduce the nonvanishing angular velocity of ZAMOs.  
In particular, along an angular direction $\phi_{j}$, with $j$ being either $1$ or $2$, the angular velocity for a zero angular momentum observer is $\Omega_{j}=g^{t\phi_{j}}/g^{tt}=-g_{t\phi_{j}}/g_{\phi_{j}\phi_{j}}$, see \cite{schutz2009first}. In this expression, $g_{x_{i}x_{j}}$ denotes the metric coefficient corresponding to the coordinates $x_{i}$ and $x_{j}$. Then with the help of \ref{eq:metric_S2_S2_SNUT} we find that $g_{t\phi_{j}}=-\Delta\,P_{j}/\rho^2$ and $g_{\phi_{j}\phi_{j}}=(\rho^2/3)\sin^2{\theta_{j}}-\Delta\,P^{2}_{j}/\rho^2$. Thus, one can obtain the explicit expression for the velocity of ZAMO along $\phi_{j}$ as
\begin{eqnarray}\label{eq:vel-ZAMO}
    \Omega_{j} &=& \frac{3\,\Delta\,P_{j}}{\rho^{4}\,\sin^2{\theta_{j}}-3\,\Delta\,P_{j}^2}~.
\end{eqnarray}
The above angular velocity vanishes when $\Delta=0$, i.e., on the horizons. Therefore, even if there is angular dragging in the spacetime, there is none on the horizons. This expression of angular velocity for a ZAMO is relevant in understanding the temperature due to Hawking radiation, see \cite{Frolov:2002xf}. However, as we shall see, the Hawking effect only concerns the angular velocity at the horizon, which vanishes in this scenario. Thus, it is expected that the Hawking temperature will not be affected by this velocity.

\begin{figure*}
    \centering
    \includegraphics[width=0.42\linewidth]{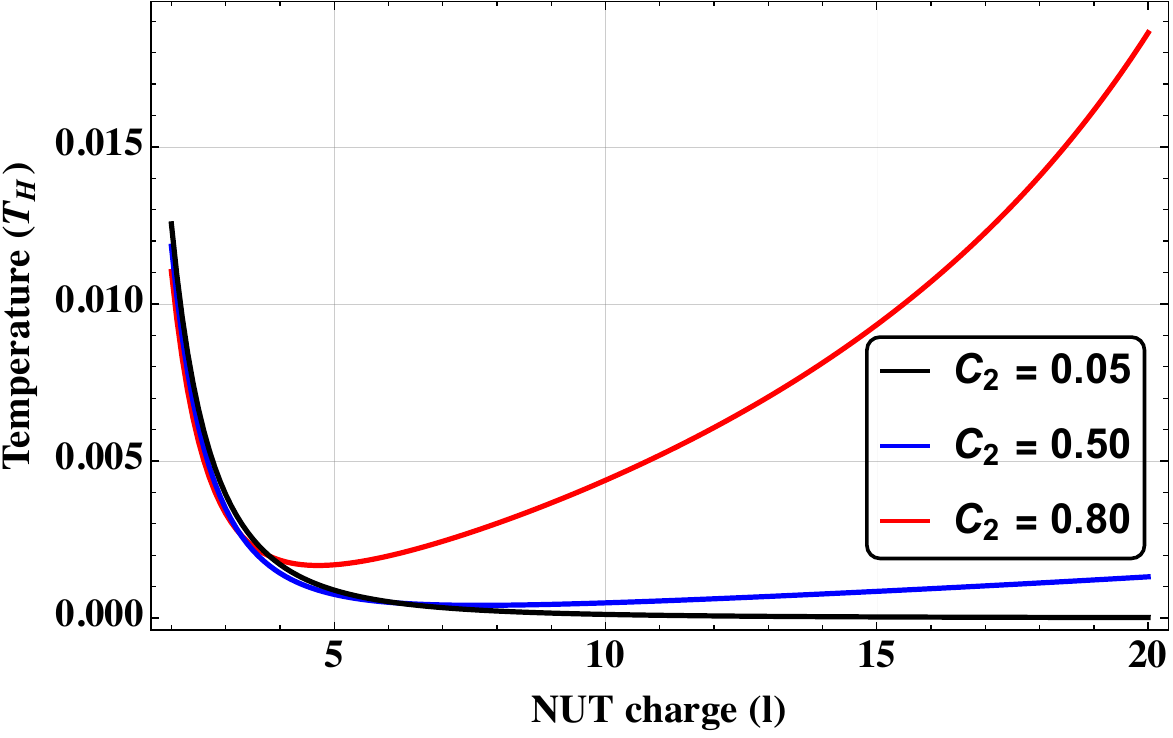}
    \hskip 20pt
    \includegraphics[width=0.42\linewidth]{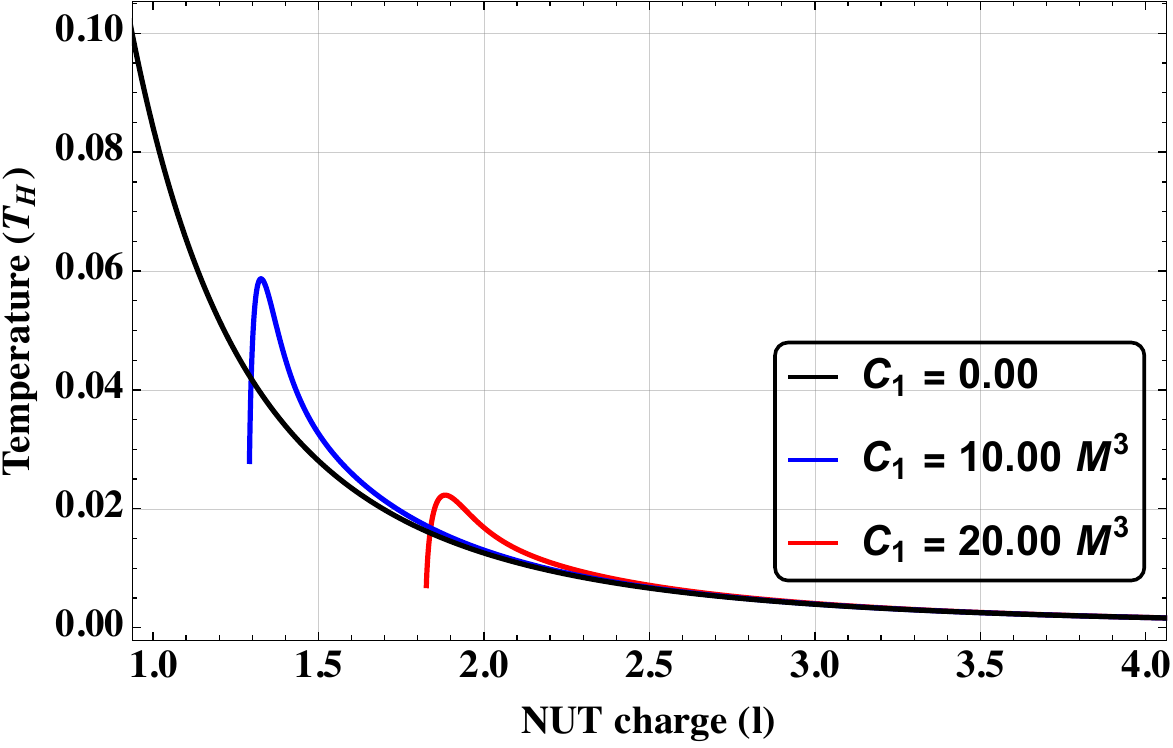}
    \vskip 10pt
    \includegraphics[width=0.42\linewidth]{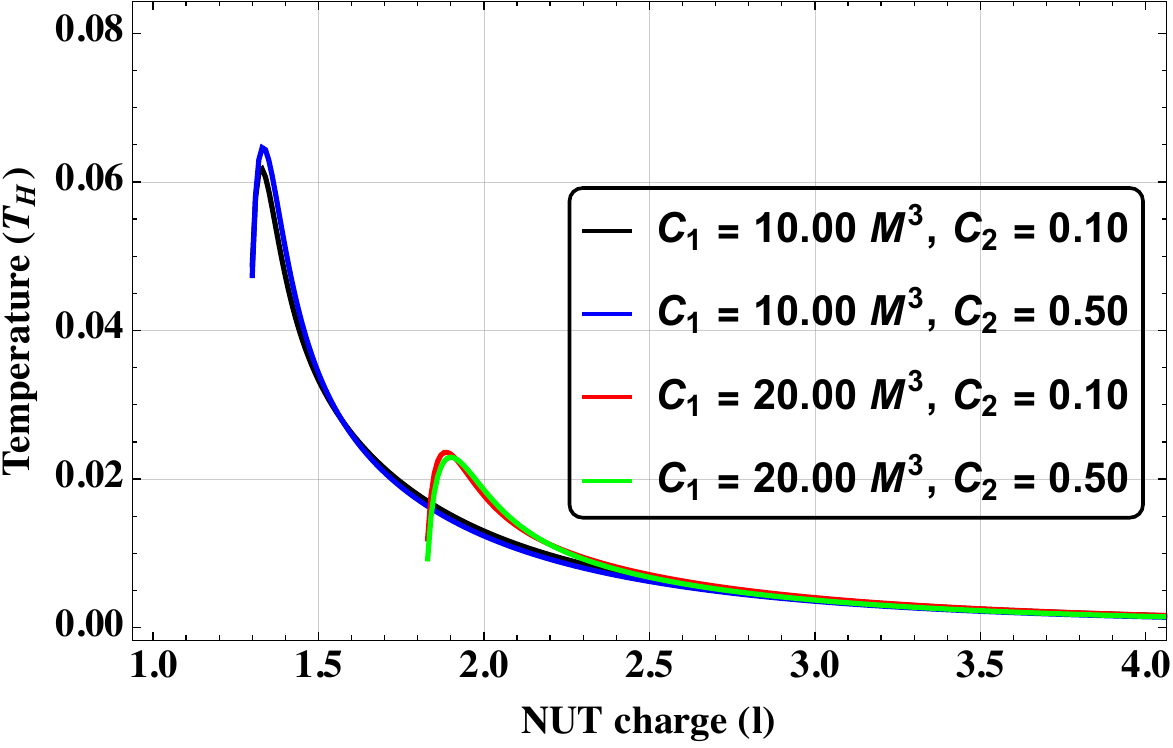}
    \hskip 20pt
    \includegraphics[width=0.42\linewidth]{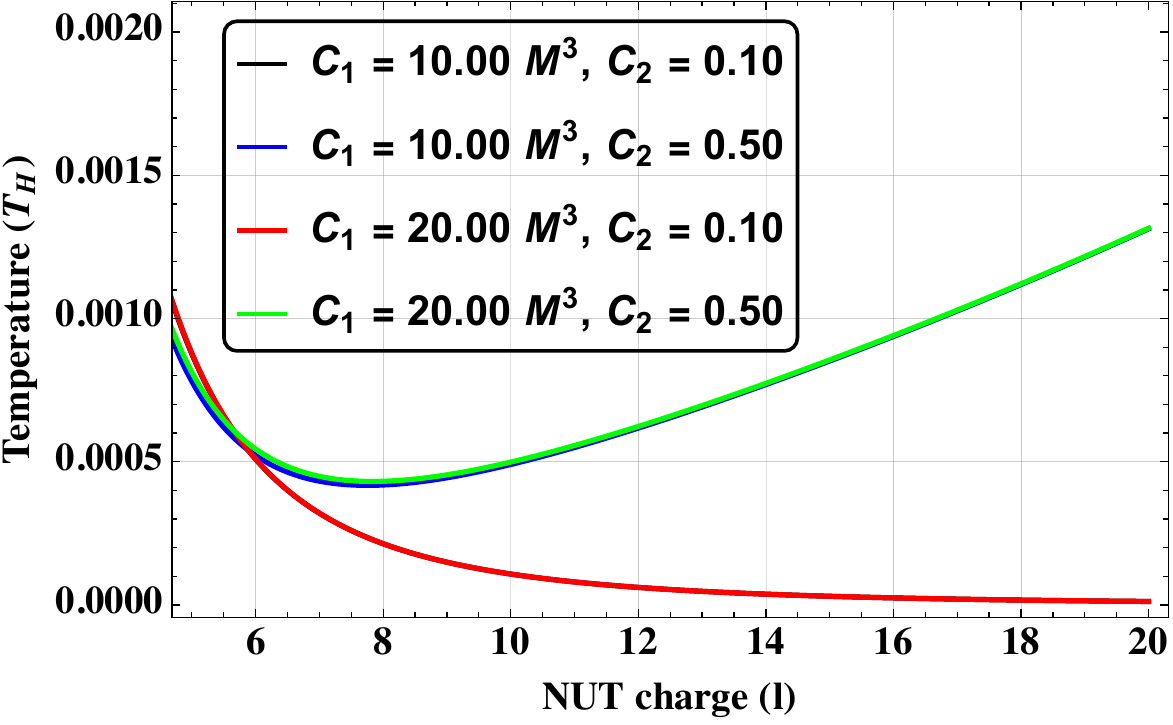}
    \caption{The above plot shows the Hawking temperature as a function of NUT charge. In the top left, we set $C_1=0$, and consider different values of $C_2$. Due to the coupling between $C_2$ and $l$, the asymptotic behavior is diverging. On the right side, we have plotted $C_2=0$ case, where we vary $C_1$. The temperature has a peak, and the location of the peak shifts towards the left if we decrease $C_1$. In the bottom plots, we have made both $C_1$ and $C_2$ to be non-vanishing and displayed both the low (left figure) and high (right figure) ends of the NUT charge. As can be seen, both the peak and tail behavior appear at the same time.}
    \label{fig:Hawking-temp}
\end{figure*}
\subsection{Thermal behavior of the spacetime}
Let us now talk about the thermal behavior of the spacetime. It is known that any spacetime with dragging on the horizon will contain the effects due to superradiance in its Hawking radiation spectra \cite{Frolov:2002xf, Ma2009HawkingTO, Jusufi:2016hcn, Barman:2018ina, Barman:2021gcd}. However, as in the current scenario, there is no drag on the horizons, one can assert that the effect due to superradiance will be absent. Then, by following the procedure described in \cite{Frolov:2002xf, Ma2009HawkingTO, Jusufi:2016hcn}, one can obtain the temperature corresponding to the Hawking effect as 
\begin{eqnarray}\label{eq:TH-Gen-HD}
    T_{H} = \frac{\kappa(r_{+})}{2\pi} &=& \lim_{r\to r_{+}} \bigg(\frac{\partial_{r}\sqrt{-G_{tt}}}{2\pi\sqrt{g_{rr}}}\bigg)~.
\end{eqnarray}
Where, $G_{tt}=g_{tt}+\sum_{j} \Omega_{j}^{+}(2\,g_{t\phi_{j}}+\Omega^{+}_{j}\,g_{\phi_{j}\phi_{j}})+\sum_{j>k}2\,\Omega_{j}^{+}\,\Omega_{k}^{+}\,g_{\phi_{j}\phi_{k}}$, with $\Omega_{j}^{+}$ being the angular velocity on the event horizon along the angular direction $\phi_{j}$, and $g_{x_{i}x_{j}}$ denote the metric coefficients corresponding to the line-element of \ref{eq:metric_S2_S2_SNUT}. As in our current scenario, the angular velocity on the event horizon along all directions is zero, we have $G_{tt}=g_{tt}$, and the temperature due to the Hawking effect becomes
\begin{eqnarray}\label{eq:TH-GBN-1}
    T_{H} = \lim_{r\to r_{+}} \bigg(\frac{\partial_{r}\sqrt{-g_{tt}}}{2\pi\sqrt{g_{rr}}}\bigg) = -\frac{1}{4\pi}\,\frac{f'(r_{+})}{r_{+}^2+l^2}~.
\end{eqnarray}

We would also like to mention that the surface gravity of the event horizon in the current scenario can be evaluated in a more straightforward manner. For instance, in this spacetime, one can always obtain a Killing vector null on the event horizon and timelike outside to be
\begin{eqnarray}\label{eq:killing-GBN}
    \xi^{\mu} = {t}^{\mu}+\sum_{j}\Omega_{j}^{+}\,\phi_{j}^{\mu}=t^{\mu}~.
\end{eqnarray}
We have the norm of this Killing vector to be $\xi_{\mu}\xi^{\mu} =g_{tt}$, which vanishes on the horizon, see \ref{eq:metric_S2_S2_SNUT}. The surface gravity $\kappa$ is obtained from the relation, see \cite{poisson2004relativist}, 
\begin{eqnarray}\label{eq:kH-GBN-1}
    2\,\kappa\,\xi_{\mu} &=& \nabla_{\mu}\big(-\xi_{\nu}\,\xi^{\nu}\big)~.
\end{eqnarray}
For the current scenario, $\xi_{\mu}=\partial_{\mu}r$ on the horizon. We also have $\nabla_{\mu}\big(-\xi_{\nu}\,\xi^{\nu}\big)=-\partial_{r}g_{tt}\, \partial_{\mu}r$. Then, the previous expression gives the surface gravity at the horizon to be
\begin{eqnarray}\label{eq:kH-GBN-2}
    \kappa(r_{+}) = \lim_{r\to r_{+}} \bigg(-\frac{\partial_{r}g_{tt}}{2}\bigg) = -\frac{1}{2}\,\frac{f'(r_{+})}{r_{+}^2+l^2}~.
\end{eqnarray}
One can notice that this expression for the surface gravity is the same as the one from \ref{eq:TH-GBN-1} obtained from the prescription of \cite{Frolov:2002xf, Ma2009HawkingTO, Jusufi:2016hcn}. If one recalls, this expression has a somewhat similar functional form in comparison to the Kerr black hole scenario, at least in the expression of the denominator. Compared to the Kerr scenario, the angular momentum per unit mass is now replaced by the NUT charge $l$. However, it should also be noted that $f'(r_{+})$ appearing in the numerator has a much more complicated expression and may not have any resemblance to the Kerr black hole.

In \ref{fig:Hawking-temp}, we have plotted the Hawking temperature corresponding to this spacetime as a function of the NUT charge $l$. We have considered three scenarios to witness the distinguishing effects caused by the coupling parameters $C_{1}$ and $C_{2}$. In one scenario $C_{1}=0$, and in the other $C_{2}=0$, we also consider a case where neither is zero. When $C_{2}$ is nonzero, we observe a change in the qualitative behavior of the temperature, especially in the large $l$ regime. It turns out that $C_2$ and the NUT charge are coupled, which gives rise to a diverging term as we increase $l$ for $C_2\neq 0$. This is visible by simply examining the top left figure in \ref{fig:Hawking-temp}. Similarly, for $C_2=0$, we can spot a peak in $T_H$ when plotting it for different $C_1$ values. With a suitable choice of non-vanishing $C_1$ and $C_2$, we capture both the peak and tail behavior, which is shown in the lower panel of \ref{fig:Hawking-temp}.

\section{Discussions} \label{sec:Discussion}
In the present article, we obtained an analytical NUT BH solution in the presence of an electromagnetic field, whereas the field equations are derived using second-order Lovelock or Gauss-Bonnet gravity. We considered a 6-dimensional spacetime and assumed the horizon topology to be $S^{(2)} \times S^{(2)}$, which is referred to as the product of two 2-spheres. Unlike the typical non-GR computation, where the deviation from \gr~appears to be a correction, in the present context, we picked up a pure Lovelock term which gives a BH solution completely different from \gr. Here, we have been primarily concerned with the interplay between NUT and Maxwell charges and how that is affecting the spacetime structure.

The major mathematical computation of the paper is obtained in \ref{sec:obtain_metric}, where we solved the field equations with the electromagnetic stress-energy tensor. We provided an exact analytical solution of the field equations and discussed the existing limiting cases in the literature. As already mentioned in Paper-I and in Ref. \cite{Pons:2014oya}, the obtained solution in pure-Lovelock in product topology is often plagued with non-central singularity, making the solution invalid in certain parameters. We have discussed these issues in \ref{sec:Validity}. It turns out that in some cases, the singularity is covered with the event horizon, and the BH solution is regular everywhere outside the horizon. However, in some cases, the singularity is located outside the event horizon, and the BH solution is not valid entirely outside the horizon. We provided a detailed discussion on the validity of the present solution by classifying it into three categories: when either of the Maxwell charges is zero and when both are present. This suggests that the qualitative structure remains the same even if the quantitative behavior alters.

The asymptotic behavior of the present BH solution is particularly interesting. Out of two Maxwell charges, $C_1$ and $C_2$, it turns out that $C_1$ plays no role in shaping the asymptotic structure. On the contrary, $C_2$ modifies the asymptotic behavior due to its coupling with the NUT charge. For easy reference, we can check \ref{eq:f(r)_S2XS2_r_infinity}, and start with the limit where both the Maxwell charges vanish. The solution does not exist unless we introduce a $\Lambda$ in the field equation. However, in the presence of an electromagnetic field, the asymptotic solution gets a dominant contribution from $C_2$, which relaxes the $\Lambda$ dependence as long as $C^2_2 l^2>27 M^2$ is valid. Interestingly, the non-existence of $C_1$ in the asymptotic expansion also hints that $C_1$ is likely to be electric, and $C_2$ is non-electric in nature.

In our investigation of horizon thermodynamics, we observed that even if the spacetime has rotation, there is no superradiance effect in the Hawking spectra. This is because the angular velocity of a ZAMO vanishes on the horizons, and a non-zero angular velocity of the horizon is what gets reflected in the superradiance effect. This result is in stark contrast to the Kerr/Kerr-Newman metric, as we observe the effects due to the superradiance phenomenon in the latter scenarios. When we compare the Hawking temperatures of the current background and a Kerr black hole, we observe that the denominators, see \ref{eq:TH-GBN-1}, have a similar form with the NUT charge now replacing the Kerr angular momentum per unit mass \cite{poisson2004relativist}. We also plotted the Hawking temperatures in the current scenario for different values of $C_{1}$ and $C_{2}$ in \ref{fig:Hawking-temp}. From this figure, we observed that the non-zero value of $C_{1}$ results in the occurrence of a peak in the Hawking temperature as one varies the NUT charge. At the same time, for non-zero $C_{2}$, the tail behavior of the Hawking temperature vs. the NUT charge curves gets significantly affected. It is to be noted that with both non-zero $C_{1}$ and $C_{2}$, there will be both the peaks and the tail features. One can also notice that the asymptotic $(r\to \infty)$ behavior of the potential from \ref{eq:pot_Maxwell_Charge} is determined by the Maxwell charge $C_{2}$. Then, it becomes natural to believe that in asymptotic regions and for large NUT charge, the Hawking temperature is relevant to analyze the role of non-electric Maxwell charge $C_{2}$ in the concerned background.

Finally, we emphasize that the present spacetime captures some interesting features emerging from the interplay between NUT and Maxwell charges. As an outcome, semi-classical observables like the Hawking temperature are affected. However, the asymptotic correction resulting from $C_2-l$ coupling may be a by-product of the chosen product topology, and it would be interesting to go further in this direction. This would also hint at how the horizon topology may impact the thermodynamical properties of a given geometry.

\section*{Acknowledgement}

The authors are thankful to Naresh Dadhich for helpful discussions at the beginning of the project. One of us (S.M.) is thankful to the Inspire Faculty Grant (DST/INSPIRE/04/2020/001332) from DST, Govt. of India, and the New Faculty Seed Grant (NFSG/PIL/2023/P3794) provided by BITS Pilani (Pilani), India, for financial support. S.B. would like to thank the Science and Engineering Research Board (SERB), Government of India (GoI), for supporting this work through the National Post Doctoral Fellowship (N-PDF, File number: PDF/2022/000428).
\appendix

\section{The tetrad components}\label{Appn:tetrad}
\label{app:tetrad}
In this section, we have written the components of the tetrad that we used in the paper. 
\begin{eqnarray}
e^{(0)}_{\mu} & = & \sqrt{\dfrac{\Delta}{\rho^2}}\left(1,0,0,0,P_1,P_2 \right), \nonumber \\
e^{(1)}_{\mu} & = & \sqrt{\dfrac{\rho^2}{\Delta}}\left(0,1,0,0,0,0 \right) \nonumber, \\
e^{(2)}_{\mu} & = & \sqrt{\dfrac{\rho^2}{3}}\left(0,0,1,0,0,0 \right)  \nonumber, \\
e^{(3)}_{\mu} & = &  \sqrt{\dfrac{\rho^2 }{3}}\left(0,0,0,\sin\theta_1,0,0 \right)  \nonumber, \\
e^{(4)}_{\mu} & = &  \sqrt{\dfrac{\rho^2}{3}}\left(0,0,0,0,1,0 \right) 
\nonumber, \\
e^{(5)}_{\mu} & = &  \sqrt{\dfrac{\rho^2 }{3}}\left(0,0,0,0,0,\sin\theta_2 \right), 
\label{eq:Tetrad}
\end{eqnarray}
where, the expressions for $P_1$, $P_2$, and $\rho$ are given below \ref{eq:metric_S2_S2_SNUT}.
\bibliographystyle{utphys1.bst}
\bibliography{GB_NUTNotes.bib}
\end{document}